\documentclass[twocolumn,twocolappendix]{aastex701}
\usepackage{amsmath}

\begin{document}

\title{Black hole astrometric binaries in the \emph{Roman} Galactic Bulge Time Domain Survey}

\author[orcid=0000-0002-6406-1924]{Casey Y. Lam}
\altaffiliation{NHFP Einstein Fellow}
\affiliation{Observatories of the Carnegie Institution for Science, 813 Santa Barbara St., Pasadena, CA 91101, USA}
\email[show]{clam@carnegiescience.edu}

\author[orcid=0000-0001-9611-0009]{Jessica R. Lu}
\affiliation{Department of Astronomy, University of California, Berkeley, CA 94720, USA}
\email[hide]{jlu.astro@berkeley.edu }  

\author[orcid=0000-0002-6871-1752]{Kareem El-Badry}
\affiliation{Department of Astronomy, California Institute of Technology, 1216 E. California Blvd., Pasadena, CA 91125, USA}
\email[hide]{kelbadry@caltech.edu}

\author[orcid=0000-0003-0248-5470]{Anil Seth}
\affiliation{Department of Physics and Astronomy, University of Utah, Salt Lake City, UT 84112, USA}
\email[hide]{aseth@astro.utah.edu}

\author[orcid=0000-0002-4733-4994]{Joshua D. Simon}
\affiliation{Observatories of the Carnegie Institution for Science, 813 Santa Barbara St., Pasadena, CA 91101, USA}
\email[hide]{jsimon@carnegiescience.edu}

\author[orcid=0000-0002-0287-3783]{Natasha S. Abrams}
\affiliation{Department of Astronomy, University of California, Berkeley, CA 94720, USA}
\email[hide]{nsabrams@berkeley.edu}

\author[orcid=0000-0003-3858-637X]{Andrea Bellini}
\affiliation{Space Telescope Science Institute, 3700 San Martin Drive, Baltimore, MD 21218, USA}
\email[hide]{bellini@stsci.edu}

\author[orcid=0000-0003-2874-1196]{Matthew W. Hosek Jr.}
\affiliation{UCLA Department of Physics and Astronomy, Los Angeles, CA 90095, USA}
\email[hide]{mwhosek@astro.ucla.edu}

\author[orcid=0000-0002-1052-6749]{Peter McGill}
\affiliation{Space Science Institute, Lawrence Livermore National Laboratory, 7000 East Avenue, Livermore, CA 94550, USA }
\email[hide]{mcgill5@llnl.gov}  

\begin{abstract}

The Nancy Grace Roman Space Telescope (\emph{Roman}), NASA's newest flagship mission, launched 30 August 2026.
As part of its mission, \emph{Roman} will conduct the Galactic Bulge Time Domain Survey (GBTDS), which will generate $\sim 50,000$ epochs of high-precision photometric and astrometric data for $\sim 10^8$ sources across 1.7 deg$^2$ in the Galactic Bulge. 
\emph{Roman} GBTDS astrometry is comparable to \emph{Gaia} Data Release 4 in terms of number of stars, astrometric precision, and  time baseline, and is highly complementary in terms of wavelength and sky location.
In this paper, we investigate a synthetic population of GBTDS sources to characterize the detectability of unresolved astrometric binaries, in particular those with compact object companions.
Assuming the occurrence rate of black holes (BHs) and neutron stars (NSs) in AU-scale orbits around stars is $10^{-7}$ and $10^{-6}$, respectively, and that \emph{Roman} achieves an astrometric precision of 1\% of a pixel, $\mathcal{O}(10)$ BH+star and $\mathcal{O}(10)$ NS+star detached binaries will be detectable.
The BHs will have median mass measurement uncertainties of $\sim 25\%$, increasing the existing sample of detached astrometric BH binaries by a factor of three.
Together with the $\mathcal{O}(10^2)$ isolated BHs expected to be discovered by microlensing in the \emph{Roman} GBTDS and an additional $\mathcal{O}(10)$ detached BH binaries in \emph{Gaia} DR4, this will provide a representative view of the quiescent Galactic stellar-mass BH population.

\end{abstract}

\section{Introduction}

The Nancy Grace Roman Space Telescope, launched 30 August 2026, will conduct surveys to uncover the nature of dark energy, map the distribution of dark matter and characterize the demographics of cold exoplanets, as well as other astrophysical investigations defined by the community \citep{Spergel:2015}.
\emph{Roman}'s primary mirror is 2.4 m, and the facility has imaging, spectroscopic, and coronagraphic capabilities in the optical and near infrared.
\emph{Roman} is optimized as a survey facility; its Wide Field Instrument (WFI) has a 0.28 deg$^2$ field of view, $100\times$ ($200 \times$) larger than that of \emph{JWST} NIRCam (\emph{HST} WFC3 IR), and its slew and settle times will be $10\times$ faster than \emph{JWST} \citep{RDox}.

Although not typically billed as an astrometry mission, the performance requirements \emph{Roman} must meet to answer questions in cosmology and exoplanets by conducting weak lensing and time-domain surveys also make it a powerful astrometric facility.
Forecasts of \emph{Roman}'s astrometric capabilities were first explored in \cite{Sanderson:2019}, with updated forecasts performed using detailed optical path simulations presented in \cite{McKinnon:2026}.
\cite{Sanderson:2019} describes several Galactic and Local Group science cases that will be enabled by \emph{Roman}'s precise astrometry, which include constraining dark matter via the orbits of the Milky Way's dwarf satellites and halo stars, finding isolated compact objects with gravitational microlensing, and understanding the puzzle of multiple stellar populations in globular clusters using proper motions.

In this work, we consider astrometric binary star orbits, whose major advantage over orbits derived from radial velocities alone is that the astrometry constrains the orbital inclination, enabling mass measurements.
Astrometric orbits have been used to measure the masses of objects in binary systems ranging from planets \citep{Holl:2023, Arenou:2023, Stefansson:2025, Sozzetti:2023} to the supermassive black hole at the center of the Galaxy \citep{Schodel:2002, Schodel:2003, Ghez:2005}.

We are particularly interested in stellar-mass black hole (BH) + star binaries.
The distribution of the BH masses, binary mass ratios, orbital periods and eccentricities reflect the outcomes of massive stellar evolution and compact object formation mechanism(s).
To date, the Galactic BH mass function \citep[e.g.,][]{Bailyn:1998, Ozel:2010, Farr:2011} and kick velocity distribution \citep[e.g.,][]{Mandel:2016, Repetto:2017} have primarily been inferred from X-ray binaries.
Although the majority of known Galactic BHs are found in X-ray binaries, there are expected to be many more BHs in wide non-interacting binaries.
A representative sample of BHs is needed to determine whether inferences made from the X-ray binary population (e.g., the existence of a mass gap between the most massive neutron stars (NSs) at $\sim 2M_\odot$ and the least massive BHs at $\sim 5 M_\odot$) are due to selection effects or if they actually reflect the physics of massive star evolution and compact object formation.

Recently, three stellar-mass BHs in wide ($1 - 20$ AU separation) binaries were identified and astrometrically characterized with \emph{Gaia} \citep{El-Badry:2023a, El-Badry:2023b, Chakrabarti:2023, Tanikawa:2023, Panuzzo:2024}.
These discoveries challenge theories of isolated binary evolution, and significant effort has been dedicated to explaining how such systems can be formed.
In addition, a $4.5 M_\odot$ BH in a wide 30 AU orbit was discovered in $\omega$ Centauri with \emph{HST} astrometry; this low-mass BH is theoretically puzzling, as the low metallicity of the cluster should produce more massive BHs \citep{Whitaker:2026}.
The discovery of additional wide BH+star binaries is necessary to characterize the population and confirm emerging relationships connecting the mass and natal kick of the BH to the binary's orbital period, eccentricity, metallicity, and environment.

In this paper, we examine the prospects for identifying and characterizing binary star systems where the individual stars are unresolved, in particular wide detached star + compact object binaries, with the \emph{Roman} Galactic Bulge Time Domain Survey (GBTDS).
We use this particular science case to explore and motivate the potential of \emph{Roman} astrometry.
Section \ref{sec:The Roman Space Telescope as an Astrometry Mission} reviews the structure of the \emph{Roman} GBTDS and its astrometric capabilities.
Section \ref{sec:Identification of binary stars} outlines how binary stars can be identified astrometrically.
Section \ref{sec:Simulations} describes the population synthesis simulations and the procedure for creating synthetic \emph{Roman} GBTDS observations.
Section \ref{sec:Results} presents the yields of detectable and characterizable binaries.
Section \ref{sec:Discussion} provides some comparison and lessons from other astrometric missions, discusses potential sources of yet unaccounted for systematic error, and investigates the resources needed for spectroscopic follow-up of candidates.
Section \ref{sec:Conclusions} presents the conclusions.

\section{The Roman Space Telescope as an Astrometry Mission
\label{sec:The Roman Space Telescope as an Astrometry Mission}}

We first provide a brief summary of the planned properties of the GBTDS.
The information in this section comes from \cite{RDox}, accessed in 2026 March\footnote{\url{https://roman-docs.stsci.edu/roman-community-defined-surveys/galactic-bulge-time-domain-survey}}.

The GBTDS fields consists of 6 \emph{Roman} WFI pointings, with 5 contiguous pointings centered at $(l, b) \approx (0.5, -1.4)$ deg, and a pointing at the Galactic Center (Figure \ref{fig:gbtds_fields}).
These 6 pointings are observed sequentially, and together compose a single GBTDS mosaic.
Each pointing has an exposure time of 66 seconds; to observe a single mosaic of all 6 pointings including slews, overheads, and exposures will take 12.1 minutes.
The mosaics will primarily\footnote{A series of ``snapshot" observations will be taken of the mosaic using all the other Roman filters as well as the grism.
As these snapshots will consist of less than 1\% of the total GBTDS observing time and contribute minimally to astrometry, we do not describe them here.} be observed in the wide F146 filter (0.93 -- 2 $\mu$m, centered at 1.46 $\mu$m), supplemented with observations in the blue F087 (0.76 -- 0.98 $\mu$m, centered at 0.87 $\mu$m) and red F213 (1.95 -- 2.30 $\mu$m, centered at 2.13 $\mu$m) filters.
The GBTDS will employ a dither strategy that includes dithers of several pixels as well as sub-pixel dithers; this results in a well-sampled pixel phase and enables precise differential astrometry across the entire WFI detector.

The Galactic Bulge is visible from \emph{Roman} twice a year for about 72 days at a time; these visibility windows are centered on the spring and autumn equinoxes.
The GBTDS consists of 3 seasons with high-cadence observations, followed by 4 seasons of low-cadence observations, followed by another 3 high-cadence seasons.
The high cadence seasons consist of continuous observations of the mosaic observed with the following filter sequence: $12\times$F146, $1\times$F087, $12\times$F146, $1\times$F213.
This results in approximately 8000 (330, 330) exposures of each field in F146 (F087, F213) during each high cadence season.\footnote{These numbers are slightly different from what is listed in \cite{RDox}.
The core community surveys and their overhead times are still being refined, so small changes are still possible.}
The low cadence seasons consist of the mosaic observed every 5 days with the following filter pattern: $5 \times$F146, $1 \times$F087, $1 \times$F213.
This results in approximately 70 (14, 14) exposures of each field in F146 (F087, F213) during each low cadence season.
The observing strategy is summarized in Figure \ref{fig:gbtds_schematic}.

\begin{deluxetable}{c|c|c}
\tablecaption{Anticipated astrometric performance of \emph{Gaia} DR4 vs. \emph{Roman} GBTDS}
\label{tab:gaia_vs_roman_ast}
\tablehead{\colhead{} & 
    \colhead{\emph{Gaia} DR4} &
    \colhead{\emph{Roman} GBTDS}} 
\startdata
Time baseline (yr) & 5.5 & 4.7 \\
\hline
Number of epochs & 40 (20--60) & 420 \\
\hline
$N_*$, $\sigma_{ast,eff}$ $\leq 0.1$ mas & 3.7 $\times 10^7$ & $>3.3 \times 10^7$ \\
 & (G $ < 15$) & (F146$_{\textrm{Vega}} < 20.6$) \\
 \hline
$N_*$, $\sigma_{ast,eff}$ $\leq 0.3$ mas & 2.2 $\times 10^8$ & $>7.5 \times 10^7$ \\
 & 2.2 (G $ < 17.5$) & (F146$_{\textrm{Vega}} < 22.1$) \\
 \hline
$N_*$, $\sigma_{ast,eff}$ $\leq 1$ mas & 7.9 $\times 10^8$ & $>1.7 \times 10^8$ \\
 & (G $ < 19.5$) & (F146$_{\textrm{Vega}} < 23.5$) \\
\hline
\enddata
\tablecomments{Time baseline, number of epochs, and number of stars with an effective per-epoch astrometric precision $\sigma_{ast,eff}$ better than (0.1, 0.3, 1) mas and the corresponding magnitude limits.
We define ``effective epochs" to more easily compare the \emph{Roman} GBTDS observations to \emph{Gaia} DR4.
We bin the \emph{Roman} observations to 1 day (as we are primarily interested in astrometric variability with timescales $\gg 1$ day), which means the number of effective epochs is smaller by a factor of $\sim 100\times$ and the effective astrometric precision is improved by a factor of $\sim 10 \times$ as compared to not binning.
See Appendix \ref{app:Assumptions about Roman GBTDS and Gaia DR4 performace} for details on how $\sigma_{ast,eff}$ and the number of \emph{Gaia} DR4 effective epochs were defined.
The number of \emph{Roman} sources is a lower limit because they are only for the 5 contiguous GBTDS field; the Galactic Center field is not included here due to limitations in available simulations of this region.}
\end{deluxetable}

Table \ref{tab:gaia_vs_roman_ast} compares the anticipated astrometric performances of \emph{Gaia} DR4 and the \emph{Roman} GBTDS.
The GBTDS surpasses \emph{Gaia} in number of epochs and has a comparable number of sources with very precise ($<0.1$ mas) epoch astrometry.
We estimate \emph{Gaia} DR4 performance based on \emph{Gaia} DR3 observations; \emph{Roman} performance was based on a simulated population and noise model described in Section \ref{sec:Simulations}.
Appendix \ref{app:Assumptions about Roman GBTDS and Gaia DR4 performace} details how these numbers were calculated.
A broader discussion of \emph{Roman} vs. \emph{Gaia} astrometry is presented in Section \ref{sec:Comparison to Gaia}.
Even before going into the detailed calculations, Table \ref{tab:gaia_vs_roman_ast} demonstrates that \emph{Roman} can and should be considered an astrometric mission.

\section{Identification of binary stars \label{sec:Identification of binary stars}}

We first define the framework for identifying astrometrically interesting sources in the \emph{Roman} GBTDS, after which we consider the specific case of binary stars.
This is modeled after the approach taken in \emph{Gaia} DR3 \citep{Arenou:2023, Halbwachs:2023}, where only sources with poor single-star fits are considered for more complex modeling.
See also \cite{Penoyre:2020, Penoyre:2022} for theoretical discussion and simulations in the context of \emph{Gaia}.

\subsection{Single-star model}

The motion of a single source (``1s") on the sky as a function of time $t$ is given by
\begin{align}
    \alpha_{*,1s}(t) &= \alpha_{*,0} + \Delta \alpha_* + \mu_{\alpha*} (t - T_p) + \varpi P_{\alpha *} (t, \alpha, \delta) \\
    \delta_{1s} (t) &= \delta_0 + \Delta \delta + \mu_\delta (t - T_p) + \varpi P_\delta (t, \alpha, \delta),
    \label{eq:single_motion}
\end{align}
where $\alpha$, $\delta$ are right ascension and declination, with $\alpha_* = \alpha \cos \delta$ and ($\alpha_{*,0}$, $\delta_0$) being a fixed reference point at which the tangent plane projection is performed.
There are 5 free parameters in the single star motion model: positional offset $(\Delta \alpha_*, \Delta \delta)$ relative to ($\alpha_{*,0}$, $\delta_0$), parallax $\varpi$, and proper motion $(\mu_{\alpha*}, \mu_\delta)$.
The parallax vector $(P_{\alpha*}, P_\delta)$ is a function of position on sky $(\alpha, \delta)$ and the observation time $t$.
$T_p$ is a reference time, which we take to be the time at periastron in the case of a binary (Section \ref{sec:Binary star model}).

\subsection{Binary star model \label{sec:Binary star model}}

With a spatial resolution of 115 mas in the F146 filter, \emph{Roman} will not resolve the individual components of the vast majority of binary systems (Section \ref{sec:Population synthesis}).
Instead, it will measure the motion of the center of light (i.e., centroid, photocenter) of the unresolved system.
The angular semimajor axis of the photocenter orbit is $a_0 = a \delta_{ql}$, where $a$ is the binary's angular semimajor axis (i.e., the semimajor axis in physical units, divided by the distance of the binary to the observer), and
\begin{equation}
    \delta_{ql} = \left| \frac{q}{q+1} - \frac{l}{l+1} \right|
\end{equation}
is the photocenter scale factor \citep{vandeKamp:1975, Belokurov:2020}.
The luminosity ratio $l = L_2/L_1$ and mass ratio $q = M_2/M_1$ are defined such that the primary (component 1) is brighter, but not necessarily more massive. 
For example, twin binaries will not be detectable because their photocenter is stationary.
But for a given period, BH + low-mass star binaries are optimally detectable, because the large mass ratio and dark companion maximizes the photocenter semimajor axis.

The orbital motion (photocenter wobble) of a binary on the sky ($\alpha_{*,O}(t), \delta_{*,O}(t)$) is given by
\begin{align}
    \alpha_{*,O}(t) &= BX(t) + GY(t) \\
    \delta_O(t) &= AX(t) + FY(t),
\end{align}
where $A, B, F, G$ are the geometric Thiele-Innes elements
\begin{align}
    A &= a_0 [\cos \Omega \cos \omega - \sin \Omega \sin \omega \cos i] \\
    B &= a_0 [\sin \Omega \cos \omega + \cos \Omega \sin \omega \cos i] \\
    F &= a_0 [-\cos \Omega \sin \omega - \sin \Omega \cos \omega \cos i] \\
    G &= a_0 [-\sin \Omega \sin \omega + \cos \Omega \cos \omega \cos i]
\end{align}
where $a_0$ is the photocenter orbit semimajor axis, $\omega$ is the argument of periapse,  $\Omega$ is the longitude of ascending node, and $i$ is the inclination.
$X, Y$ are the dynamical Thiele-Innes elements
\begin{align}
    X(t) &= \cos E(t) - e \\
    Y(t) &= \sqrt{1 - e^2} \sin E(t)
\end{align}
where $e$ is the orbital eccentricity and $E(t)$ is the eccentric anomaly \citep{Green:1985}.
$E(t)$ is solved for in Kepler's equation
\begin{equation}
    E - e \sin E = \frac{2\pi}{P_{orb}} (t - T_p)
\end{equation}
where $P_{orb}$ is the orbital period.
Then the complete motion of a binary system (``2s") on the sky is the sum of its center-of-mass motion (described by single star motion) and the photocenter wobble
\begin{align}
    \alpha_{*,2s}(t) &= \alpha_{*,1s}(t) + \alpha_{*,O}(t) \\
    \delta_{2s}(t) &= \delta_{1s}(t) + \delta_O(t).
    \label{eq:binary_motion}
\end{align}

\subsection{Identifying binaries in Roman 
\label{sec:Identifying binaries in Roman}}

The astrometric goodness-of-fit statistic to a single-star model $\chi^2_{1s}$ is defined
\begin{equation}
    \chi^2_{1s} = \sum_i^N \left[ \left( \frac{\alpha_{\textrm{model},i} - \alpha_{\textrm{data},i}}{\sigma_i} \right)^2 + \left( \frac{\delta_{\textrm{model},i} - \delta_{\textrm{data,i}}}{\sigma_i} \right)^2 \right]
\end{equation}
where $(\alpha_{\textrm{model},i}, \delta_{\textrm{model},i})$ are the predicted model positions, $(\alpha_{\textrm{data},i}, \delta_{\textrm{data},i})$ are the measured positions, $\sigma_i$ are the measured positional uncertainties, and $N$ is the number of epochs of observations; we assume the positional uncertainties are the same in both directions, and that all epochs are independent measurements.
For a single source model, the number of degrees of freedom is $k = 2N - 5$.

In the case where $k \gg 1$, the distribution of $\chi^2_{1s}$ follows a normal distribution with mean $k$ and standard deviation $\sqrt{2k}$.
For ease of interpretation, we transform our $\chi^2_{1s}$ values to follow the standard normal distribution.
The standard score for the single-star model fit $z_{1s}$ is defined to be
\begin{equation}
    z_{1s} = \frac{\chi^2_{1s} - k}{\sqrt{2k}}
    \label{eq:z}
\end{equation}
and follows a standard normal distribution.
Thus, large values of $z_{1s}$ (e.g., $z_{1s} > +5$) are indicative of single source models being poor descriptions of the observations.

One way to identify binary systems in the \emph{Roman} GBTDS is to model all sources with a 5-parameter single-star model, identify which sources are poorly modeled (i.e., have large $z_{1s}$), and then re-model the source with a 12-parameter binary-star model to determine whether an orbit is a better description of the data.
This is the approach taken by the \emph{Gaia} mission in DR3 \citep{Halbwachs:2023}.
The main questions on applying this to \emph{Roman} are how to define the threshold for a ``poor" fit and how to balance completeness (i.e., including as many binaries as possible) against purity (i.e., number of false positive sources that have poor single-star fits but are not actually binary sources).
Although these questions can only be definitively answered once actual \emph{Roman} data are in hand, we discuss lessons learned from \emph{Gaia} in Section \ref{sec:Binary contaminants}.

\section{Simulations \label{sec:Simulations}}

All magnitudes listed in this work are in the Vega system.
Photometry information provided by the \emph{Roman} project in their documentation is given in AB magnitudes.
We use the conversion $F146_{\mathrm{AB}} = F146_{\mathrm{Vega}} + 1.012$ \citep{Lancaster:2022}.

\begin{figure}
    \centering
    \includegraphics[width=1.0\linewidth]{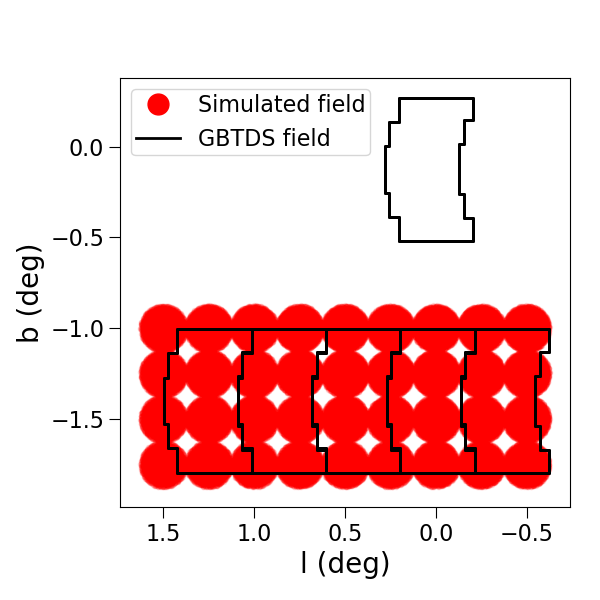}
    \caption{
    \emph{Roman} GBTDS footprint (black outline) with the locations of the simulated fields run in this paper (red circles).
    We only simulate observations within the five contiguous GBTDS fields; we do not simulate observations toward the GBTDS Galactic Center field.
    Note that the GBTDS autumn mosaic configuration is shown; the spring mosaic configuration has the GBTDS pointings rotated by 180 degrees.}
    \label{fig:gbtds_fields}
\end{figure}

\begin{figure}
    \centering
    \includegraphics[width=1.0\linewidth]{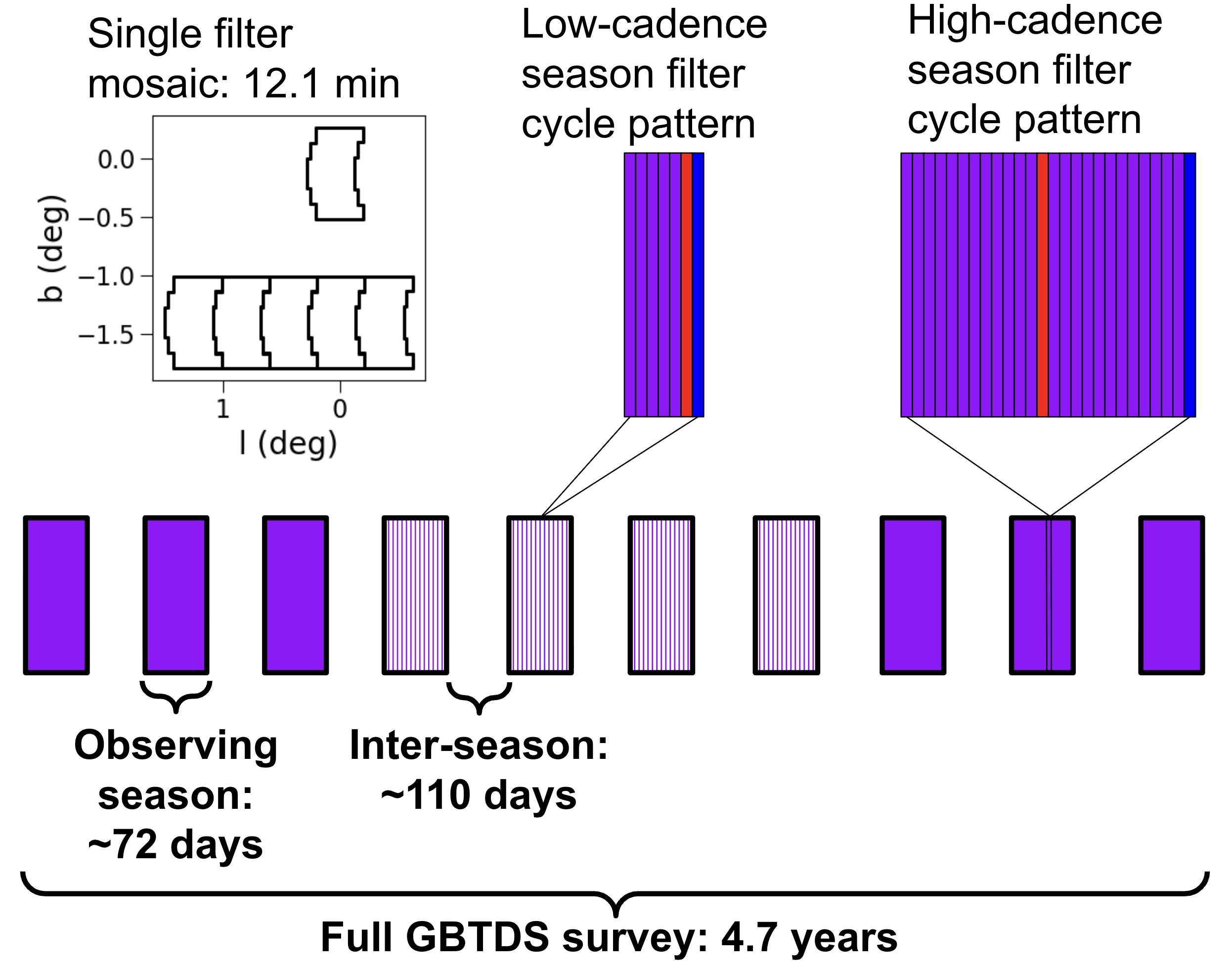}
    \caption{Schematic of the observational structure of the GBTDS.
    Six fields toward the Galactic Bulge are observed sequentially in a cycle; this is a GBTDS mosaic.
    During high-cadence observing seasons (first three and last three seasons), this mosaic is observed continuously; during low-cadence observing seasons (middle four seasons), this mosaic is observed roughly every 5 days.
    The filters in which this mosaic is observed are schematically shown; blue is F087, purple is F146, and red is F213.
    Note that the temporal axis is not drawn to scale.
    }
    \label{fig:gbtds_schematic}
\end{figure}

\subsection{Population synthesis \label{sec:Population synthesis}}

We perform Galactic population synthesis to create a mock source catalog with positions, velocities, magnitudes, and binary orbital properties.
In this section and the rest of the paper, we define a single source to be a bound stellar system; thus, a single star is counted as one source, and a binary star system is also counted as one source.
This will roughly correspond to a source extraction algorithm's definition of a point source; we justify this below from the results of the population synthesis.

The population synthesis is performed using \texttt{PopSyCLE} \citep{Lam:2020}.
The starting point of \texttt{PopSyCLE} is the single-star Besan\c{c}on Galactic model \citep{Robin:2003}, as implemented in the \texttt{galaxia} software package \citep{Sharma:2011}.
The bar angle and length are modified following the description in Appendix A of \cite{Lam:2020} (their ``v3" model).
In \texttt{PopSyCLE}, synthetic photometry is generated using \texttt{SPISEA} \citep{Hosek:2020}, using the isochrone and atmosphere models listed in Section 2.2 of \cite{Lam:2020}.
The extinction model uses the $E(B-V)$ color excesses from \cite{Schlegel:1998} and the \cite{Damineli:2016} reddening law.

As the Bescan\c{c}on model only includes luminous stars, \texttt{PopSyCLE} performs population synthesis of white dwarfs, NSs, and BHs to be included in the Galactic model.
Compact objects are generated by assuming the progenitor star is replaced by a compact object following the initial-final mass relations described in \cite{Rose:2022} (their ``Spera15" model, \cite{Spera:2015}). 
\texttt{PopSyCLE} has also been modified to include binary systems as described in Section 2 of \cite{Abrams:2025}, where the orbital period, mass ratio, and eccentricity distributions are drawn following the distributions presented in \cite{Duchene:2013, Kobulnicky:2007, Kiminki:2012, Lu:2013} and the multiplicity statistics are from \cite{Lu:2013}. 

The population synthesis in \cite{Abrams:2025} does not include the effects of binary evolution, interaction, or destruction; although this is a good assumption for main sequence binaries, it is not true for evolved systems involving compact objects.
In order to generate a more realistic population of compact object + star binaries, we modify the output of the compact object prescription of \texttt{PopSyCLE} as follows, using simple prescriptions that are observationally motivated.

For binaries with a white dwarf component, we follow the prescription used in \cite{El-Badry:2024b, Lam:2025, Yamaguchi:2025}, which assumes the following:
\begin{itemize}
    \item When the white dwarf progenitor is in the red giant or asymptotic giant branch phase, unstable mass transfer results in a common envelope evolution phase that tightens some, but not all, of the binaries with orbital semimajor axis $< 6$ AU down to $P_{orb} \lesssim 1$ day \citep{Shahaf:2024, Yamaguchi:2024} \footnote{It would be better to account for eccentricity and make this a function of periastron distance, but we ignore this for our simplified prescription.}.
    We thus assign 10\% of WD + star binaries with initial separations of 2 -- 6 AU to have current separations that are 50\% of their initial separations.
    The rest of the systems with separations $< 6$ AU are assigned final orbital periods of $P_{orb} = 1$ day, which are not detectable astrometrically.
    \item For binaries with initial separations $> 6$ AU, we assume that the orbits change only due to mass loss, ignoring other potential types of interactions such as wind mass transfer.
    When mass loss occurs on a timescale longer than the orbital period, the product of  total mass times semimajor axis is conserved \citep{Jeans:1924}, so that
    \begin{equation}
        a_i = a_f \frac{M_f + M_*}{M_i + M_*}
    \end{equation}
    where $a_i (a_f) $ is the initial (final) semimajor axis, $M_i$ ($M_f$) is the WD's progenitor (current) mass, and $M_*$ is the companion star's mass.
    The relationship between $M_f = 0.109 M_i + 0.394M_\odot$ is from \cite{Kalirai:2008}.
\end{itemize} 

The \emph{Gaia}-measured occurrence rate of NS and BH binaries in AU-scale orbits is $\mathcal{O}(10^{-6})$ per star and $\mathcal{O}(10^{-7})$ per star, respectively \citep{Lam:2025, Nagarajan:2025, El-Badry:2024a}.
\texttt{PopSyCLE} by default produces a factor of $\sim 20$ NSs and $\sim 100$ more BHs than the \emph{Gaia} rate implies.
Thus, we removed a random 95\% and 99\% of \texttt{PopSyCLE}-generated binaries containing NSs or BHs, respectively.
This is physically motivated by the expectation that the majority of BH and NS binaries will be disrupted during their formation due to a Blaauw and/or natal kick \citep[e.g.,][]{Wiktorowicz:2019}.
The remaining systems retain their initial orbital configurations, with the more massive star replaced by a compact object; we assume no other interactions occur.
Although this portion of the prescription is unphysical (e.g., mass loss will widen the orbit), it enables us to match the observed NS and BH occurrence rate.

The population synthesis simulations are run in a grid of circular pointings covering the contiguous GBTDS fields\footnote{
These simulations come from a larger grid spanning Galactic pointings $l \in [-0.5, 1.5]$ deg $\times$ $b \in [-1.75, 0.25]$ deg in 0.25 deg steps, and were originally generated for field selection optimization studies in the definition of the GBTDS. 
We reuse these simulations as part of this work, which is why the simulations are not centered on the finalized GBTDS fields.} (Figure \ref{fig:gbtds_fields}).
We simulate 36 pointings, each covering an area of 0.05 deg$^2$, for a total area of 1.8 deg$^2$.
Since the total area of the 5 contiguous GBTDS fields is only $0.281 
\times 5 = 1.405$ deg$^2$, we rescale our simulated numbers of sources by a factor of 0.78.

\begin{figure*}
    \centering
    \includegraphics[width=\linewidth]{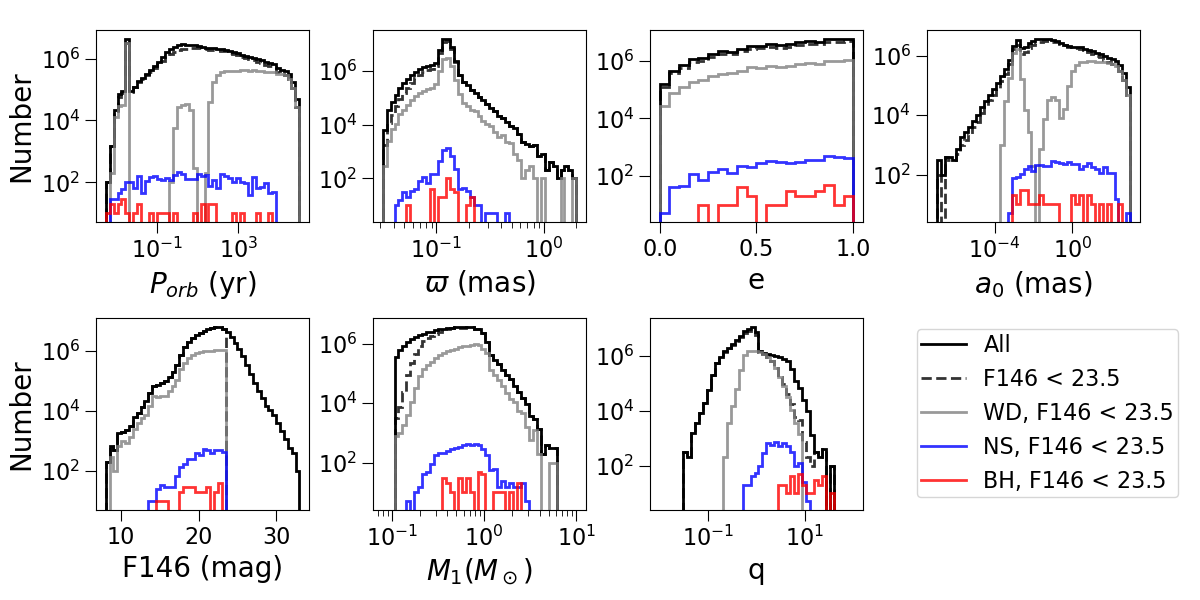}
    \caption{Simulated luminous binary population properties (orbital period $P_{orb}$, parallax $\varpi$, eccentricity $e$, angular photocenter orbit semimajor axis $a_0$, F146 magnitude, primary mass $M_1$, and mass ratio $q$) for the contiguous GBTDS fields.
    The solid black line denotes the distribution for all sources within 30 kpc.
    The dashed black line denotes the subset ($\sim 80\%$) which are F146$_{\textrm{Vega}} < 23.5$ mag (source dominated regime).
    The gray, blue, and red outlines denote the population of white dwarf, NS, and BH binaries, respectively, with F146$_{\textrm{Vega}} < 23.5$ mag.
    Note that the primary is defined to be the brighter component of the binary, and the mass ratio is defined to be the ratio of the secondary to the primary; this can lead to $q > 1$, especially in the case of BH+star binaries.
    }
    \label{fig:contiguous_field_binaries}
\end{figure*}

Figure \ref{fig:contiguous_field_binaries} shows the properties of the simulated binary population.
After rescaling, there are $1.6 \times 10^8$ sources with F146$_{\textrm{Vega}} < 23.5$ mag ($4.7 \times 10^5$ sources will be saturated with F146$_{\textrm{Vega}} \lesssim 15$ mag). 
In the bright F146$_{\textrm{Vega}} < 23.5$ mag subset of stellar systems, 1/3 will be binary systems. 
We note this is not the intrinsic binary fraction, which is much lower ($\sim 10\%)$ due to low-mass stars being the most common, and having a very low binary fraction.
However, because most of these stars are not observable due to their faintness, the binary fraction of detectable bright stars, which are more massive, is higher.

For simplicity in interpretation, we define resolved binaries to be systems where the angular semimajor axis of the binary is larger than the full-width half-maximum (FWHM) of the Roman point-spread function (PSF) in the F146 filter, which is 115 mas \citep{McKinnon:2026}.
The vast majority (99\%) of binary sources F146$_{\textrm{Vega}} < 23.5$ mag with two luminous components will be unresolved with Roman in F146 (Figure \ref{fig:contiguous_field_binaries}) which makes our assumption at the beginning that stellar binaries are considered single sources reasonable.

\begin{figure}
    \centering
    \includegraphics[width=\linewidth]{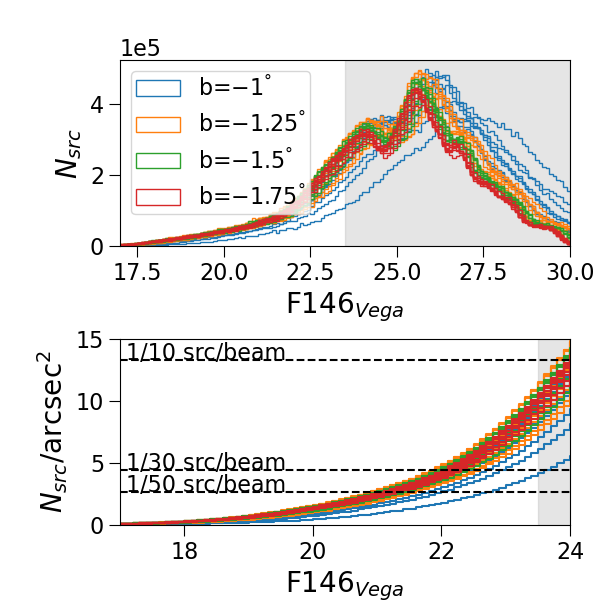}
    \caption{Top panel: simulated luminosity function (number of sources per magnitude bin) for the contiguous GBTDS fields.
    This luminosity function includes \emph{all} luminous sources, not just binaries (c.f. Figure \ref{fig:contiguous_field_binaries}).
    The field latitudes are labeled by the different colored lines; longitudes are not labeled for readability.
    The gray shaded regions F146$_{\textrm{Vega}} > 23.5$ mag denotes the background dominated regime.
    Bottom panel: Cumulative source density distribution, zoomed in on the region of optimal astrometric sensitivity.
    The horizontal dashed lines denote 1/10, 1/30, and 1/50 sources per beam, which are a proxy for the amount of crowding (discussed in Section \ref{sec:Astrometric precision and systematic errors}).
    }
    \label{fig:luminosity}
\end{figure}

The luminosity function (as a histogram as well as a cumulative density) towards the contiguous GBTDS fields is shown in Figure \ref{fig:luminosity}.
The fields are extremely crowded; this will be discussed more in Section \ref{sec:Discussion}.

Due to two major shortcomings in the current simulations, we do not create a synthetic catalog for the GBTDS Galactic Center pointing.
The first issue is that the \texttt{galaxia} simulations do not include the nuclear stellar disk (NSD) around the Galactic Center.
The NSD has a physical radius of 230 pc and scale height of 45 pc \citep{Launhardt:2002};  at the distance of the Galactic Center (8 kpc) this corresponds to an angular radius of $1.6^\circ$ and scale height of $0.3^\circ$, and will span the full field field of view of the GBTDS Galactic Center pointing.
The NSD is expected to increase the number of stars by $\sim 2\times$ in the Galactic Center field.
The second issue is that the method of estimating the extinction used in \cite{Lam:2020, Abrams:2025} is inaccurate for highly extincted regions and wide wavelength filters.
These issues are actively being addressed in an updated version of these simulations (M. Huston et al. in preparation).

Given that we are neglecting one out of the six GBTDS fields, the numbers of stars, detected objects, etc. quoted throughout this work are all lower limits on the true outcome of the GBTDS. 
Even if the number of stars in the Galactic Center field is higher by an order of magnitude than one of the contiguous GBTDS fields, this will only underestimate the total number of GBTDS sources by a factor of 3.
Given the other major uncertainties (e.g., impact of crowding, patchy extinction, astrometric performance), this is acceptable for this paper.
We note that the effects of ignoring the NSD and the issues with the extinction estimation are minimal for the five contiguous GBTDS fields, because they are at least several scale heights away from the NSD, and the extinction is significantly less away from $b=0$ deg.

\subsection{Mock Roman catalogs}

Next, we create mock \emph{Roman} star catalogs.
The GBTDS survey and configurations are based on the Roman User Documentation \citep{RDox} as accessed in March 2026.

\subsubsection{Temporal Sampling \label{sec:Temporal Sampling}}

Since we are interested in astrometry, we only simulate the observations taken in the F146 filter.
For simplicity, we assume all fields have 70 days of continuous observations during the high cadence seasons, with observations taken every 12.1 minutes. 
Every 13th observation is excluded as it is in the F087 or F213 filter.
For the low cadence seasons, we assume the seasons are 70 days, with a set of observations taken every 5 days, where each set of observations consists of 5 exposures taken every 12.1 minutes.
This results in a total of about 48,000 epochs taken in F146.

\subsubsection{Noise Model \label{sec:Noise Model}}

\begin{figure}
    \centering
    \includegraphics[width=1.0\linewidth]{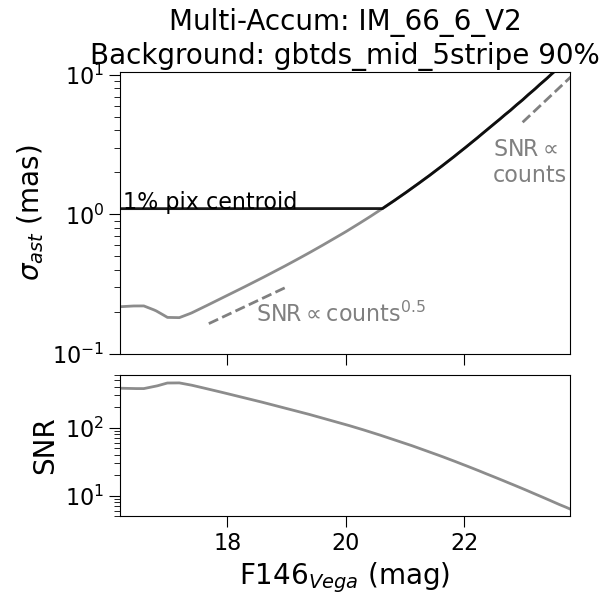}
    \caption{GBTDS F146 per-exposure astrometric precision $\sigma_{ast}$ (top panel; solid gray line) and SNR (bottom) as a function of magnitude calculated from the simulations of \cite{McKinnon:2026}.
    The dashed lines show the scaling relationship in the source dominated regime (SNR $\propto$ counts$^{0.5}$) and the background dominated regime (SNR $\propto$ counts).
    Centroiding of 1\% of a pixel is a commonly assumed astrometric precision \citep{Sanderson:2019, McKinnon:2026}, corresponding to a floor of 1.1 mas (black curve) for \emph{Roman}.}
    Centroiding to $\sim$0.1\% of a pixel would be required to take full advantage of the SNR in the source dominated regime.
    \label{fig:roman_ast_noise_model}
\end{figure}

Our noise model comes from the \emph{Roman} astrometry simulation tool of \cite{kevin_mckinnon_2026_18407082}, which is built upon the \texttt{Pandeia} simulations \citep{Pontoppidan:2016}.
\texttt{Pandeia} produces a 2-D SNR map of a simulated observation for a given astronomical scene (defined by the source(s) spatial profile and spectral energy distribution, background level) and the instrumental setup (filter, exposure configuration) using a simulated super-sampled \emph{Roman} PSF from \texttt{stpsf} \citep{Perrin:2012, Perrin:2025} for the appropriate filter.
The \cite{kevin_mckinnon_2026_18407082} tool simulates a point source using \texttt{Pandeia}, takes the generated 2-D SNR map and iteratively measures the flux and position of the source. 
We note that these models are highly idealized; for example, issues due to crowding and confusion are also not included in these simulations, which may be an important systematic (Section \ref{sec:Discussion}).

We run the \cite{kevin_mckinnon_2026_18407082} \emph{Roman} astrometry simulation following the information provided in \cite{RDox} for the GBTDS.
We use the multi-accumulation table IM\_66\_6\_V2, corresponding to a 66 second total exposure and 5 up-the-ramp resultants\footnote{A resultant is a group of averaged detector reads.} and 1 reset/read resultant.
We use the background ``gbtds\_mid\_5stripe", which is calculated at the center of the 5 contiguous fields.
To be conservative, we use the ``high" level background, corresponding to the 90th percentile of the background distribution.
Figure \ref{fig:roman_ast_noise_model} shows the per-exposure astrometric precision\footnote{In this work, when we say ``centroiding precision" or ``astrometric precision", we mean the 1D uncertainties in the $x$ or $y$ direction.
We assume the uncertainties in $x$ and $y$ are the same.} $\sigma_{ast}$ and SNR as a function of F146$_{\textrm{Vega}}$ magnitude for a single GBTDS exposure.

The astrometric precision for time series observations not only depends on the single exposure centroiding precision, but on other systematics (Section \ref{sec:Astrometric precision and systematic errors}).
Unless otherwise stated, we impose a 1\% centroiding precision in the analysis going forward; for \emph{Roman}'s 0.11 arcsec pixels this corresponds to a precision of 1.1 mas per exposure.
This impacts bright sources F146$_{\textrm{Vega}} < 20.62$ mag (Figure \ref{fig:roman_ast_noise_model}).
The 1\% centroiding precision was chosen as this level of astrometric precision has been demonstrated with HST \citep{Bellini:2011,Kallivayalil:2013,Hosek:2015}.
As the centroiding precision for \emph{Roman} will not be known until GBTDS observations have been collected over time, we choose this fairly conservative assumption.

We also emphasize that this 1\% centroiding floor is added as white noise.
This implicitly assumes that each observation is an independent measurement and that the measurement errors are all uncorrelated.
To achieve this in reality will require systematic errors (including time dependent systematic errors that will vary across the full 5 year temporal baseline) to be sub-dominant to the per-epoch measurement errors; see Section \ref{sec:Astrometric precision and systematic errors} for discussion on the sources of systematic errors.

\subsubsection{Final catalog}

For each source in the simulated catalog, we calculated orbital tracks using Equation \ref{eq:binary_motion} at all the expected observing times in F146 as described in Section \ref{sec:Temporal Sampling}.
The positions were then perturbed using the Gaussian noise prescription in Section \ref{sec:Noise Model} and then each point was assigned the appropriate per-epoch astrometric error according to the brightness of the star.
We re-emphasize this fundamentally assumes all measurements are independent and measurement errors are uncorrelated.

\section{Results \label{sec:Results}}

We focus on sources with F146$_{\textrm{Vega}} < 23.5$ mag, which corresponds to $\sigma_{ast} \approx 10$ mas and where the SNR starts to become background dominated.
Although the GBTDS is certainly sensitive to sources fainter than F146$_{\textrm{Vega}} < 23.5$ mag, the astrometric precision rapidly degrades at these fainter magnitudes, especially in crowded regions.
To be astrometrically sensitive to orbits at worse precision would require the orbits to be larger; in most cases these would be such long periods \emph{Roman} would not constrain them well with its $\approx 5$ year time baseline.
They would also be more difficult to confirm with spectroscopic follow-up.
Thus, we focus on brighter sources with better astrometry.

\begin{figure*}
    \centering
    \includegraphics[width=0.325\linewidth]{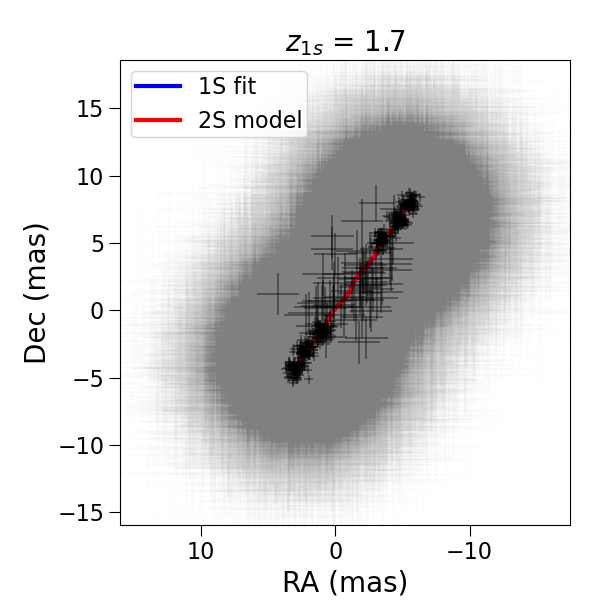}
    \includegraphics[width=0.325\linewidth]{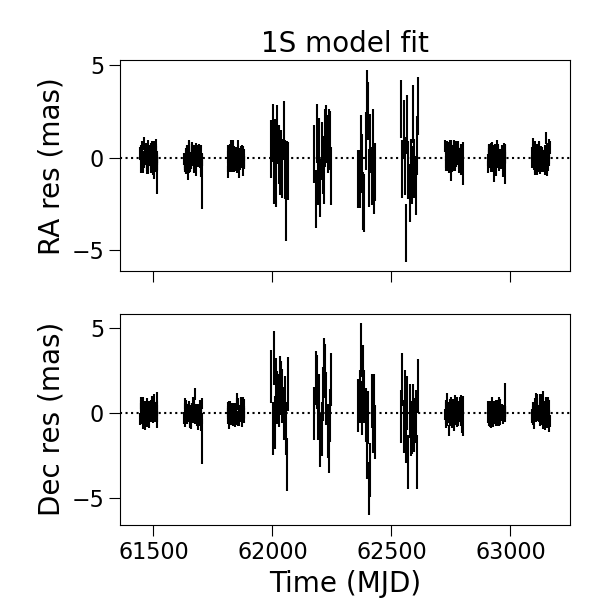}
    \includegraphics[width=0.325\linewidth]{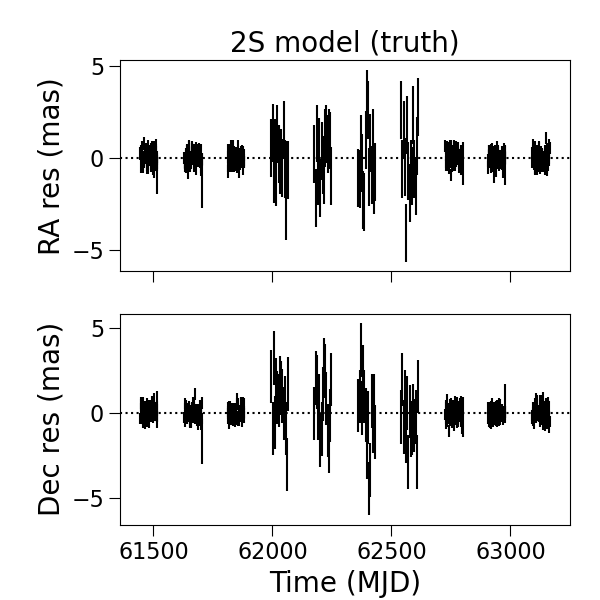}
    \includegraphics[width=0.325\linewidth]{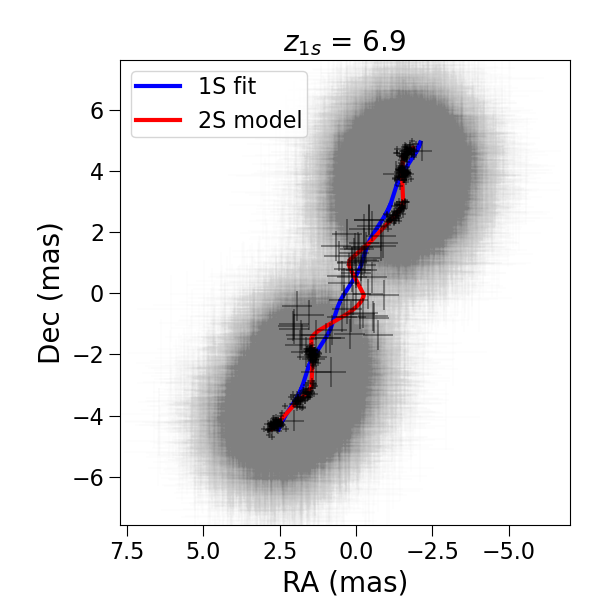}
    \includegraphics[width=0.325\linewidth]{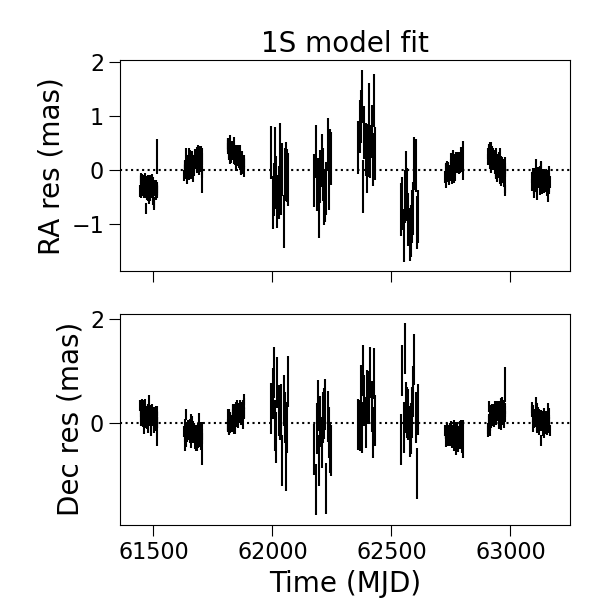}
    \includegraphics[width=0.325\linewidth]{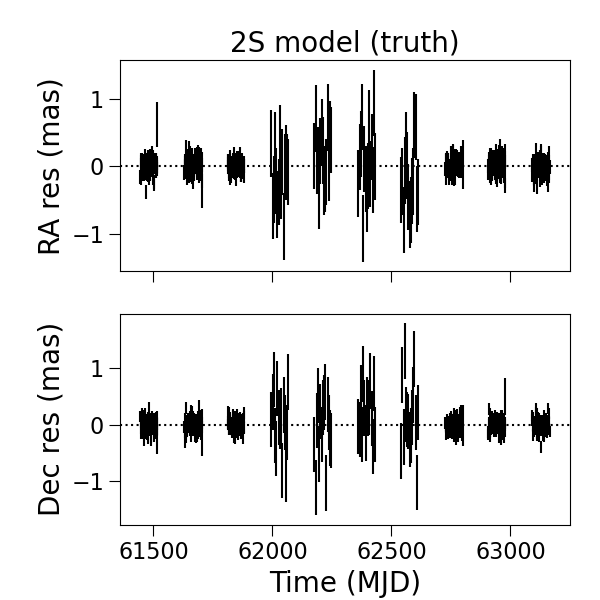}
    \includegraphics[width=0.325\linewidth]{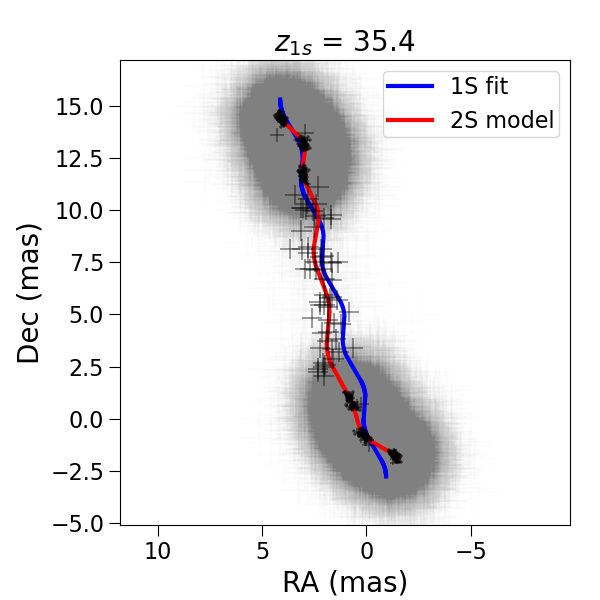}
    \includegraphics[width=0.325\linewidth]{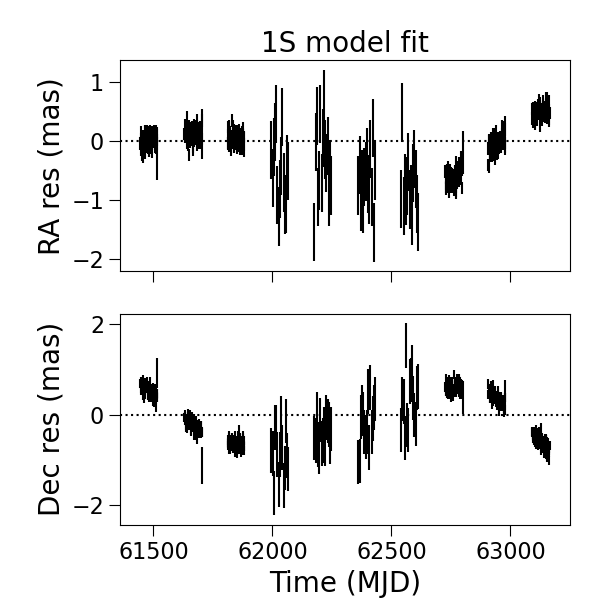}
    \includegraphics[width=0.325\linewidth]{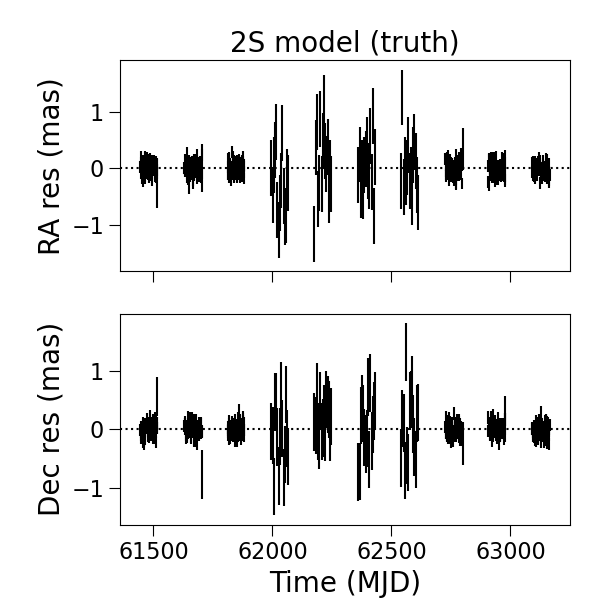}
    \caption{Simulated astrometry for three binaries.
    The left panels show the on-sky trajectories; gray points are individual epoch measurements and black points are those measurements binned to 1 day.
    The gray points separate into two ``clouds" in the middle and bottom panels due to the structure of the GBTDS, as the high-cadence seasons occur at the beginning and end of the survey, with low cadence seasons in the middle.
    The red curve traces the true orbital motion; the blue curve shows the best fit single star model.
    The middle (right) panels show the residuals to the single (binary) star model.
    Top row: A star+star binary with with $z_{1s} = 1.7$.
    Input parameters: $P_{orb} = 1251$ days, $e = 0.68$, $d = 8.2$ kpc ($\varpi = 0.12$ mas), $a = 0.02$ mas, $a_0 = 0.25$ mas, F146$_{\textrm{Vega}}$ = 22.2 mag, $M_1 = 0.44 M_\odot$, $M_2 = 0.33 M_\odot$, $\vec{\mu} = (1.9, -2.7)$ mas yr$^{-1}$.
    Middle row: A BH+star binary with $z_{1s} = 6.9$.
    Input parameters: $P_{orb} = 539$ days, $e = 0.60$, $d = 7.6$ kpc ($\varpi = 0.13$ mas), $a = 0.43$ mas, $a_0 = 0.41$ mas, F146$_{\textrm{Vega}}$ = 18.8 mag, $M_1 = 0.84 M_\odot$ (star), $M_2 = 15.1 M_\odot$ (BH), $\vec{\mu} = (1.00, -1.99)$ mas yr$^{-1}$.
    Bottom row: A BH+star binary with $z_{1s} = 35.4$.
    Input parameters: $P_{orb} = 1680$ days, $e =0.31 $, $d = 4.45$ kpc ($\varpi = 0.22$ mas), $a = 1.24$ mas, $a_0 = 1.12$ mas, F146$_{\textrm{Vega}}$ = 18.8 mag, $M_1 = 0.76 M_\odot$ (star), $M_2 = 7.1 M_\odot$ (BH), $\vec{\mu} = (-1.13, -3.56)$ mas yr$^{-1}$.
    The single star residuals in the $z_{1s} = 6.9$ and 35.4 examples demonstrate how the single-star model is a poor fit; in the $z_{1s} = 1.7$ case the single star model is an adequate description because the binary signal is not detectable given the uncertainties.
    \label{fig:z1_z6_orbit}}
\end{figure*}

\begin{figure}
    \centering
    \includegraphics[width=1.0\linewidth]{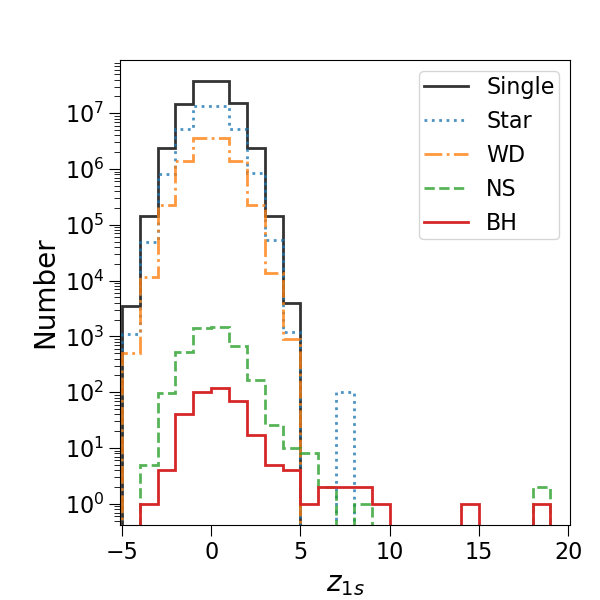}
    \caption{Distribution of the goodness-of-fit $z_{1s}$ of sources in the GBTDS contiguous field with F146$_{\textrm{Vega}} < 23.5$ mag.
    The distribution of $z_{1s}$ for luminous single sources is shown in solid black; their $z_{1s}$ distribution follows a standard normal distribution.
    The distributions of $z_{1s}$ for luminous binary sources (Star + Star, WD, NS, or BH) are shown in the dotted blue, dot-dash orange, dashed green, and solid red lines, respectively.
    BH+star binaries have a significant tail beyond $z_{1s} \gtrsim 3$, due to model mismatch.
    \label{fig:z_dist}}
\end{figure}

\begin{deluxetable}{c|c|c}[h!]
\tablecaption{Binary detectability}
\label{tab:binary detectability}
\tablehead{\colhead{Type} & 
    \colhead{$N_{src}$, F146$_{\textrm{Vega}} < 23.5$} &
    \colhead{$N_{src}$ detect (\%)}} 
\startdata
Star+Star & $3.9 \times 10^7$ & $\leq 100$ ($\leq 0.0003 $\%) \\
WD+star & $1.0 \times 10^7$ & $\leq 100$ ($\leq 0.001$\%) \\
NS+star & $4.4 \times 10^3$ & 13 (0.3\%) \\
BH+star & $3.7 \times 10^2$ & 10 (3\%) \\
\enddata
\tablecomments{Number of bright (F146$_{\textrm{Vega}} < 23.5$ mag) binaries and the subset of which are identifiable as binaries with $z_{1s} > 5$ for 1\% pixel per exposure astrometric precision broken down by binary type.
Note these are for the total number of stars in the 5 contiguous \emph{Roman} GBTDS field (1.4 deg$^2$ total); it does not include the GBTDS Galactic Center field.}
\end{deluxetable}

\begin{figure*}
    \centering
    \includegraphics[width=\linewidth]{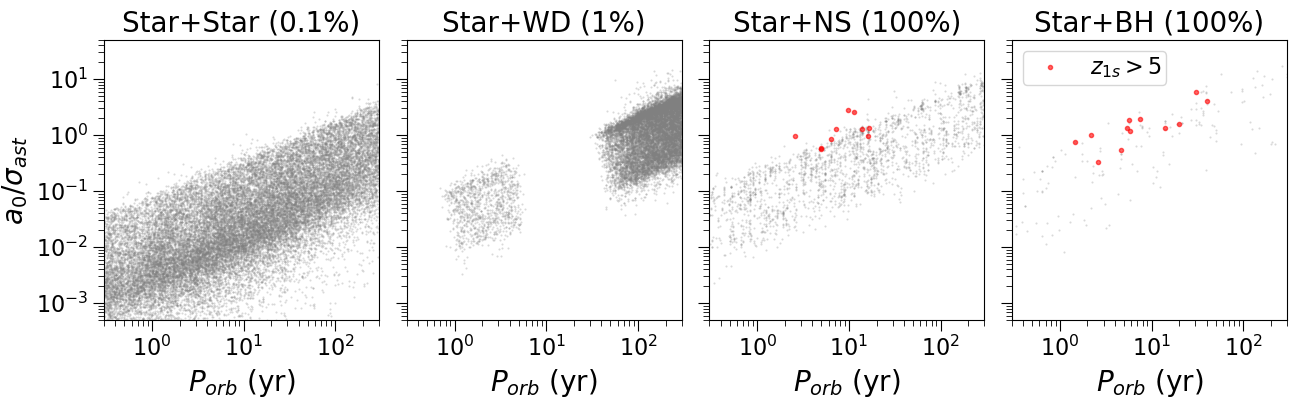}
    \caption{Detectability of binaries in the contiguous GBTDS fields, broken down binary type.
    The number in parenthesis indicates the fraction of the population that is shown in the figure; as there are many more star+star binaries than BH+star binaries, we only show a small fraction of the former for clarity.
    Gray points indicate sources with F146$_\textrm{Vega} < 23.5$ mag; red points indicate the subset of those which are astrometrically detectable ($z_{1s} > 5$).
    The positive correlation is due to Kepler's 3rd law as $P_{orb} \propto a^{3/2}$, which is then modulated by scatter in $\delta_{ql}$, $\varpi$, and $\sigma_{ast}$.
    Roman is most sensitive (i.e., large $z_{1s}$) to periods of $1 - 10$ years.
    The gap in the white dwarf distribution is due to the choice made in population synthesis (Section \ref{sec:Population synthesis}). 
    \label{fig:porb_a0_over_error_z_gc}}
\end{figure*}

\begin{figure*}
    \centering
    \includegraphics[width=1.0\linewidth]{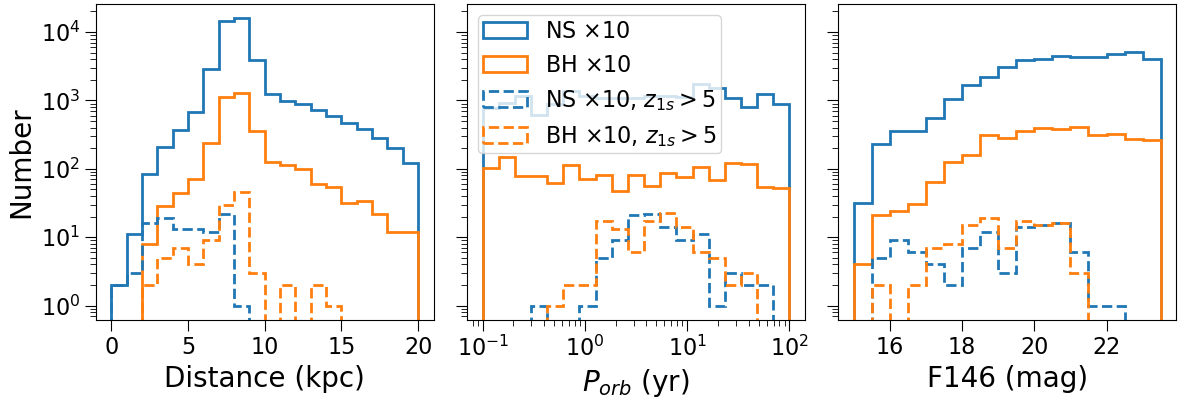}
    \caption{Distance, orbital period $P_{orb}$, and F146$_{\textrm{Vega}}$ magnitude distributions of NS+star and BH+star 
    binaries in the GBTDS contiguous fields.
    All binaries with F146$_{\textrm{Vega}} < 23.5$ mag are shown in the solid lines; binaries with poor single star goodness-of-fits ($z_{1s} > 5$) are shown in dashed lines.
    Because NSs are less massive than BHs, they have smaller photocenter wobbles and are preferentially found at closer distances.
    Note that we have generated $10\times$ as many NS and BH binaries as in our fiducial model above to enable better visualization of the statistics.
    \label{fig:distance_porb_mag}}
\end{figure*}

\subsection{Detectability of unresolved binaries}

We fit the $\approx 50,000$ epochs of simulated \emph{Roman} GBTDS astrometric binary measurements generated in Section \ref{sec:Simulations} to a single-star model (Equation \ref{eq:single_motion}), and then calculated the resulting standard score for the fit $z_{1s}$ (Equation \ref{eq:z}).

To reduce the computational cost, we only fit a random subsample of the binaries. 
We simulate $3.9 \times 10^5$ star+star binaries, $1.0 \times 10^5$ WD+star binaries, $4.4 \times 10^3$ NS+star binaries, and $3.7 \times 10^2$ BH+star binaries; this is 1\%, 1\%, 100\%, and 100\% of the total population of F146$_{\textrm{Vega}} < 23.5$ mag binaries in the contiguous fields.
Thus, the underlying population of F146$_{\textrm{Vega}} < 23.5$ mag binaries consists of $3.9 \times 10^7$ star+star binaries, $1.0 \times 10^7$ WD+star binaries, $4.4 \times 10^3$ NS+star binaries, and $3.7 \times 10^2$ BH+star binaries.

Figure \ref{fig:z1_z6_orbit} shows simulated astrometry for a star+star binary with low $z_{1s}$, a BH+star binary with moderate $z_{1s}$, and a BH+star binary with very high $z_{1s}$.
Figure \ref{fig:z_dist} shows the distribution of $z_{1s}$ for all \emph{Roman} GBTDS sources F146$_{\textrm{Vega}} < 23.5$ mag.
The tail of high $z_{1s}$ sources is primarily due to the model mismatch in fitting a single star model to BH+star and NS+star binaries.

In this work, we define a binary as ``detected" if its $z_{1s} > +5$.
This is chosen based on a false positive rate from true single stars assuming Gaussian statistics, as a 5$\sigma$ detection is fairly robust to statistical outliers; 
systematic errors are discussed in Section \ref{sec:Binary contaminants}.
In general, the detectability of a binary is a function of its orbital period (because the survey time baseline is finite), and the size of its photocenter wobble relative to the astrometric uncertainties.
For $P_{orb} \lesssim \Delta T_{survey}$, analytic expressions can be derived (i.e., Equations 17 and 20 in \cite{Penoyre:2020}) to estimate the excess astrometric noise relative to a single-source fit.

The results of binary detectability are summarized in Table \ref{tab:binary detectability}.
Out of the full population of F146$_{\textrm{Vega}} < 23.5$ mag binaries, with 1\% pixel astrometric precision we detect 0.3\% of the detached NS+star binaries and 3\% of detached BH+star binaries; scaled to the full sample of binaries this corresponds to total numbers of $\sim 10$ NS+star and $\sim 10$ BH+star binaries.
Out of our simulated star+star and WD+star population, we detect 1 and 0 systems, respectively; for Poisson errors this is consistent with zero, and so we quote upper limits on the number of detected systems.

Figure \ref{fig:porb_a0_over_error_z_gc} shows the results of the binary detectability in the plane of $a_0/\sigma_{ast}$ vs. $P_{orb}$.
Sources that are detectable tend to have orbital periods $1 \lesssim P_{orb} \lesssim 10$ years; the lower limit is set by orbit size (when it becomes too small to detect given the astrometric precision) and the upper limit is set by the survey time baseline (the portion of the orbit detected is sufficiently small such that it is only detected as linear motion).
For a given orbital period, systems with the largest $a_0/\sigma_{ast}$ are the detectable systems.
NS and BH binaries are most detectable because at a given orbital period, they have larger photocenter wobbles due to being massive and dark.

Figure \ref{fig:distance_porb_mag} further breaks down the detectable NS and BH binaries by their distances, orbital periods, and magnitudes. 
In order to visualize the statistics more easily, we have generated $10\times$ more binaries than are included in our fiducial model.
The orbital period distribution for the NS and BH binaries are similar.
The detectable NS binaries tend to be slightly closer on average than the BH binaries, as they have smaller photocenter wobbles.

\subsection{Ability to recover orbital parameters}

\begin{figure*}
    \centering
    \includegraphics[width=0.49\linewidth]{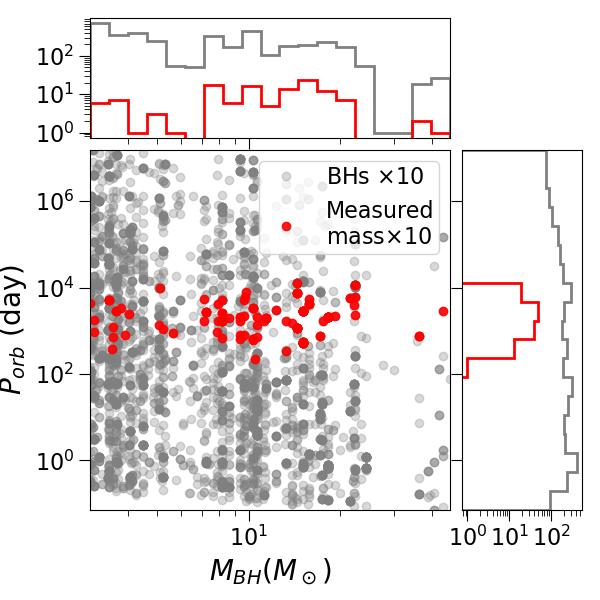}
    \includegraphics[width=0.49\linewidth]{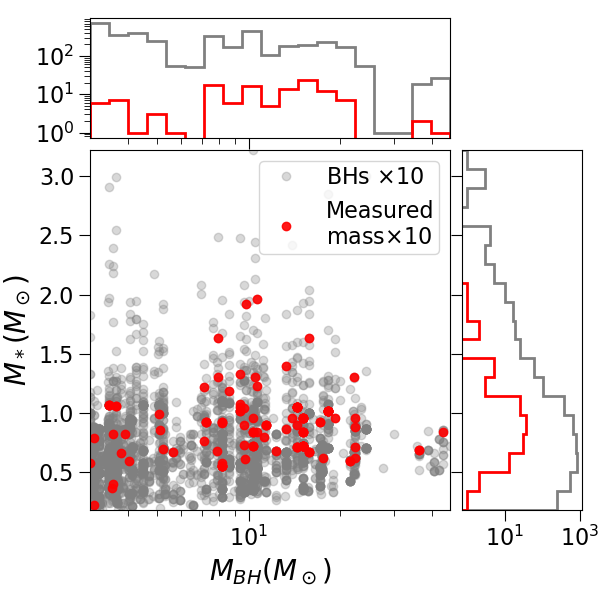}
    \caption{BH+star binaries simulated in the \emph{Roman} GBTDS contiguous fields.
    All simulated sources are shown in gray; sources for which masses are measured are shown in red.
    The left (right) panel shows their orbital periods (stellar companion mass) vs. BH masses.
    Note that we have generated $10\times$ as many NS and BH binaries as in our fiducial model above to enable better visualization of the statistics.
\label{fig:MBH_vs_F146_vs_Porb}}
\end{figure*}

Most of this paper deals with the \emph{identification} of binaries in the GBTDS.
However, the ultimate goal is \emph{characterization} of these systems, in particular compact object+star binaries.
Although a detailed study of the recovery of binary orbital parameters and population inference is out of scope of this work, we perform a simple analysis here to estimate the expected types of constraints that will be achievable on BH+star binaries. 

We modeled a sample of 125 BH+star binaries with $z_{1s} > 5$; we again simulated an extra 10$\times$ as many binaries to improve the statistics.
To speed up the model evaluation and reduce computational time, we bin the observations into 1 day stacks.

We use Markov chain Monte Carlo (MCMC) as implemented by \texttt{emcee} \citep{Foreman-Mackey:2013} to fit binary models to our mock \emph{Roman} BH+star observations.
We performed our fitting with 100 walkers, 2000 steps (of which 1000 were burn-in), and used every 10th step in the chain; we initialized our walkers approximately 10\% away from the true values, using wide flat priors on the parameters.
As a cross check on our MCMC, we used the nested sampling algorithm \citep{Skilling:2004} which is better for sampling degenerate and bumpy likelihoods, to fit binary models to a few targets; the results were consistent within errors to our MCMC fits.
Four of our fits failed to converge.
It is possible that with additional work (e.g., better priors, running longer or with more walkers) a solution could have been derived.

For the remaining BH+star binaries with reliable measurements, orbits with $P_{orb} \lesssim 5$ years have fractional period measurement uncertainties $\sigma_{P_{orb}}/P_{orb} \lesssim 1\%$, those with $P_{orb} \sim 5$ years have uncertainties $\sigma_{P_{orb}}/P_{orb} \sim \textrm{few}\%$, and those with $P_{orb} \sim 30$ years have uncertainties $\sim 20\%$.
We note what determines how well the true orbit can be recovered is primarily the orbital period, and not $z_{1s}$; $z_{1s}$ only helps prioritize which orbits should be fit with more complex models, as it would be extremely computationally expensive to fit a binary star model to every single source in the GBTDS.

The main quantities of interest are the astrometric mass function of the binary (as high mass functions are indicative of BH companions, which should be followed up) and the BH mass that can be derived.
The astrometric mass function
\begin{equation}
    f_M = \left( \frac{a_0}{\varpi} \right)^3 \left( \frac{P_{orb}}{\textrm{yr}} \right)^{-2} = (M_1 + M_2) \delta_{ql}^3 < M_1 + M_2
\end{equation}
places a lower limit on the mass of the binary.
For a dark companion, such as a BH where the primary is the star $(*)$ and the secondary is the BH $(\bullet)$, the astrometric mass function can be written as
\begin{equation}
    f_M = \frac{M_\bullet^3}{(M_\bullet + M_*)^2} .
\end{equation}
Thus, assuming a measurement of $f_M$ and $M_*$, we can directly constraint the BH mass $M_\bullet$.
The uncertainty on the measurement of $M_\bullet$ is
\begin{equation}
    \sigma_{M_\bullet} = \frac{(M_\bullet + M_*)^2 \sigma_{f_M} + 2 f_M (M_\bullet + M_*) \sigma_{M_*}}{3M_\bullet^2 - 2f_M(M_\bullet + M_*)}.
\end{equation}

The stellar companion's mass can be estimated by using GBTDS multi-band photometry covering $0.62 - 2.125 \mu$m to measure the spectral type and class of the star.
For low-mass stars (which are expected to dominate the Bulge population), this will enable $\sim 20\%$ uncertainties on the mass of the star.
For low-mass star ($\lesssim 2 M_\odot$) + BH binaries, we find typical BH mass measurement uncertainties of $\sigma_{M_\bullet} = 3.0^{+3.0}_{-1.5} M_\odot$, corresponding to mass uncertainties of $\sigma_{M_\bullet}/M_\bullet = 24^{+12}_{-6}\%$.
This is comparable to the median mass uncertainties of $\sim 25\%$ on the primary mass of merging binary BHs detected via gravitational waves \citep{LVK:2026}, which will enable a robust measurement of the detached BH binary mass function. 

Figure \ref{fig:MBH_vs_F146_vs_Porb} shows the BH+star binaries with mass measurements in orbital period--BH mass $P_{orb} - M_{BH}$ and stellar companion mass--BH mass $M_*- M_{BH}$ space.
As in Figure \ref{fig:distance_porb_mag}, we have generated $10\times$ more binaries than are included in our fiducial model for visualization purposes.
The main thing to note is that the selection criteria are not particularly sensitive to the mass of the BH; this is similar to astrometric orbits with \emph{Gaia} \citep{Lam:2025}.
Thus, the raw sample of BH masses discovered in detached binaries in \emph{Roman} will be fairly similar to the completeness-corrected population.
This also means \emph{Roman} will be sensitive to BHs in the mass gap, if they exist in the Milky Way.
However, the periods and distances will not be representative due to strong selection effects introduced by the selection criteria on $z_{1s}$.

\section{Discussion \label{sec:Discussion}}

\subsection{Comparison to Gaia \label{sec:Comparison to Gaia}}

The only other astrometric mission comparable to the \emph{Roman} GBTDS in both size and precision is \emph{Gaia} (Section \ref{sec:The Roman Space Telescope as an Astrometry Mission}, Table \ref{tab:gaia_vs_roman_ast}).
They are highly complementary with regards to observing strategy, sky location, and wavelength, which enables them to probe different parts of the Galaxy.
Here, we compare and contrast various aspects of \emph{Gaia} and \emph{Roman} GBTDS.
See the review of \cite{Brown:2021} for comparisons between \emph{Gaia} and \emph{HST} astrometry; many of the comments about \emph{HST} astrometry are also broadly applicable to \emph{Roman}.

\emph{Gaia} astrometry is absolute, due to its scanning strategy and anchoring its reference frame to fixed quasars across the sky.
\emph{Roman} GBTDS astrometry is relative and differential, where the stars themselves are used to iteratively define a reference frame; relative astrometry can be placed in an absolute frame by cross-matching common sources to a reference.

\emph{Gaia} astrometry is white-light optical ($G$, 330 -- 1050 nm), \emph{Roman} GBTDS is primarily NIR (F146, 930 -- 2000 nm).
\emph{Gaia} observes all-sky, but performs poorly in the Bulge due to crowding and extinction  \citep{Fabricius:2021, Luna:2023}.
On the other hand, \emph{Roman} GBTDS is conducted toward the crowded dusty Bulge where \emph{Gaia}'s astrometric abilities degrade.
Although \emph{Roman} GBTDS astrometry will also likely be affected by crowding at some level, it will still perform significantly better than Gaia (see the comparisons in \cite{Luna:2023} to HST and VIRAC2).
\cite{McKinnon:2026} estimate \emph{Roman} GBTDS will deliver proper motion uncertainties $\sigma_\mu < 1$ mas yr$^{-1}$ for sources $G < 29$, parallax uncertainties $\sigma_\varpi < 1$ mas for sources $G < 28$, and fractional distance uncertainties $\sigma_r/r < 10\%$ for sources $G < 22.5$ and $r < 10$ kpc; these estimates also do not account for crowding.

The forward model described in Section \ref{sec:Simulations} predicts the median distance (50\% range) of $1 M_\odot$ MS stars in Gaia with good astrometry ($G < 19.5$ mag) is 4.0 kpc (2.7 -- 5.7 kpc).
For Roman GBTDS sources with good astrometry (F146$_{\textrm{Vega}} < 23.5$ mag) this is 9.5 kpc (7.2 -- 12.3) kpc for the 5 contiguous fields.
\emph{Roman} probes distances much closer toward the center of the Galaxy than \emph{Gaia}, where the stars span a wider range of metallicities and the stellar density is higher than in the Solar neighborhood. 
Having samples of detached BHs from both \emph{Roman} and \emph{Gaia} will reveal whether or not detached BH formation strongly changes in these different environments.

The temporal sampling of the two surveys is very different. 
Due to \emph{Gaia}'s scanning law, the number of observations for a given sky position is variable, but a rough rule of thumb is that any given source will be observed at a monthly cadence.
On the other hand, the \emph{Roman} GBTDS has a much higher cadence (12 minutes or 5 days; Section \ref{sec:The Roman Space Telescope as an Astrometry Mission}) during its observing seasons, but these are interrupted by long gaps due to the restrictions on Bulge visibility; however, because the seasons are separated by 6 months, this is ideal for parallax measurements.

When performing population studies, selection functions are the crucial link needed to make inferences from individual detections. 
Homogeneous surveys with publicly accessible data make it much easier to robustly determine selection functions.
This is a strength of the \emph{Roman} GBTDS and \emph{Gaia} DR4, which will enable the statistics of binaries in the solar neighborhood from \emph{Gaia} to be compared with those discovered by \emph{Roman} toward the Bulge.

\emph{Gaia} reports several goodness-of-fit metrics to evaluate their single source model fits.
The renormalized unit weight error \texttt{ruwe} has been found to be the most reliable metric; the renormalization removes systematics introduced by calibration errors of bright, blue, and very red sources \citep{Lindegren:2021}.
In Gaia DR3, \texttt{ruwe} is directly related to the astrometric gaussianized chi-squared \texttt{astrometric\_gof\_al}\footnote{Per the Gaia DR3 documentation, 
``Goodness-of-fit statistic of the astrometric solution for the source in the along-scan direction. 
This is the ``gaussianized chi-square", which for good fits should approximately follow a normal distribution with zero mean value and unit standard deviation. Values exceeding, say, $+3$ thus indicate a bad fit to the data."}, which follows a standard normal distribution for well-behaved single sources.
Our standard score $z_{1s}$ is equivalent to \texttt{astrometric\_gof\_al}.
In the construction of the astrometric non-single star catalog in \emph{Gaia} DR3, only sources with $\texttt{ruwe} > 1.4$ were considered for fitting with a binary model \citep{Halbwachs:2023}.
In \emph{Gaia} DR3, $\texttt{ruwe} = 1.4$ is roughly equivalent to $z_{1s} \approx 5 - 12$ (depending on the number of independent observations; fewer (more) observations corresponds to a smaller (larger) $z_{1s}$).
We note that for the upcoming \emph{Gaia} DR4, this threshold for trying a binary model may change.

For the \emph{Roman} GBTDS, $z_{1s}$ will be reported as part of the public data products, so it will be simple to make a cut on this parameter.
Although we considered $z_{1s} > 5$ as the cutoff in this paper, this value was motivated based on statistical, not systematic, errors.
For \emph{Roman}, although the statistical error floor will be very low due to the sheer number of measurements, the systematic floor is unknown, and this will be important in setting what the appropriate $z_{1s}$ floor will be.
Systematic errors and false positives are discussed in more detail in the following sections.

\subsection{Comparison to HST}
Although we have primarily been comparing the \emph{Roman} astrometry to \emph{Gaia}, in many aspects \emph{HST} astrometry in crowded fields is a much closer analog.
Here, we consider some lessons learned from crowded field relative astrometry with \emph{HST}.

\subsubsection{Unresolved orbit searches}

The core of the globular cluster $\omega$ Centauri ($\omega$ Cen, located a distance 5.4 kpc from Earth, \citep{Baumgardt:2021}) is frequently observed by \emph{HST}; in addition to general observer science programs, it is the calibration source for HST WFC3/UVIS.
Because of this, there is an extensive, albeit heterogeneous, set of astrometric observations spanning a $>20$ year baseline dataset consisting of $\sim 30$ epochs.
\cite{Platais:2024} and \cite{Whitaker:2026} used these archival observations to search for unresolved astrometric binaries to look for detached BHs.
Both \cite{Platais:2024} and \cite{Whitaker:2026} identify a source with a strong acceleration signature; \cite{Platais:2024} find it is a $\sim 10$ year NS binary, while \cite{Whitaker:2026} finds it to be a $\sim 100$ year BH binary; see Appendix \ref{app:Details of the omega Cen black hole search} for a more detailed summary of the discovery and interpretation of this system.

The results of \cite{Platais:2024} and \cite{Whitaker:2026} are very promising for \emph{Roman}.
The GBTDS's temporal baseline of $\approx 5$ years will enable orbits with $P_{orb} \lesssim 12$ years to have well-constrained masses; longer periods will also have good mass measurements if the orbits are eccentric and the observations happen at periastron, as in the case of the BH in $\omega$ Cen.
Although \emph{Roman} will not have as long a time baseline and so will not be sensitive to such long orbits, and the HST observations were also shorter wavelengths (336 nm -- 814 nm) and thus have better spatial resolution than the \emph{Roman} F146 observations will, the \emph{Roman} observations will be significantly more homogeneous, span a much larger field of view, and have many more epochs.

\subsubsection{Astrometric precision and systematic errors
\label{sec:Astrometric precision and systematic errors}}

The treatment of astrometric precision in this paper was based upon the \texttt{Pandeia} simulations, which effectively gives us knowledge of the statistical uncertainties (i.e., variations due to the number of photons), which \cite{McKinnon:2026} translated into an astrometric uncertainty.
\texttt{Pandeia} does not include the effects of systematics such as crowding or geometric distortion, as it is estimated from a single point source.
Similarly, our modeling of the inferred positions only accounts for statistical errors, not systematic errors. 
We perturb the positions according to a Gaussian distribution based on the size of the uncertainties from \texttt{Pandeia}, which is a function of the source SNR.
However, we do not model the actual process of astrometric alignment, where the sources are placed onto a common reference frame, which is defined from the stars themselves.

Here, we present some estimates on the magnitude of systematic effects and the ability to correct them, based on experience from \emph{HST} and other \emph{Roman} simulations.
\emph{Roman} is very similar to \emph{HST} WFC3/IR, but with a significantly larger ($\sim 200 \times$) field of view; \emph{Roman} also has the advantage over \emph{HST} of being at L2, which will be more stable than a low Earth orbit.
See also the discussion on astrometric precision, systematic errors, how to mitigate them, and HST experience in \cite{McKinnon:2026} and \cite{Sanderson:2019}.

\emph{Crowding.}
Given the uncertainty in the stellar density and extinction, it is difficult to characterize the effect of crowding without the actual observations.
We make some simple estimates here using our simulations and comparing to observational datasets.
We quantify crowding in terms of the number of sources per beam $\theta_{beam} = \pi (\textrm{FWHM}/2.35)^2$ \citep{Hogg:2001}.
For Roman in the F146 filter, FWHM = 115 mas and thus $\theta_{beam} = 0.0075$ arcsec$^2$.
Our estimate of the number of sources per arcsec$^2$ given our simulated luminosity function is shown in Figure \ref{fig:luminosity}.

Based on HST WFC3/IR observations of the Arches and Quintuplet clusters in the near-IR, the completeness begins to sharply fall due to confusion once the source density exceeds 1/50 -- 1/30 sources/beam \citep{Hosek:2022}, corresponding to F146$_{\mathrm{Vega}} = 21 - 22$ mag.
We find that most binaries need to be this bright in order to be astrometrically detected (Figure \ref{fig:distance_porb_mag}); thus, we do not expect confusion to substantially change the number of detectable sources as predicted in this work, unless the SNR of these sources drops sharply because of high backgrounds due to unresolved stars.
However, the number of sources in Table \ref{tab:gaia_vs_roman_ast} with $\sigma_{ast} \leq1$ mas is likely overestimated.

Even for sources that are separated by several pixels, the presence of nearby sources can introduce excess astrometric scatter by a factor of $2-5$ (M. Whitaker et al. in preparation).
The severity of this will depend on the contrast of the stars and the filter; it will be worse for faint stars near bright stars, and worse at longer wavelengths.
This ``wide separation crowding" can also introduce large systematic errors that need to be corrected.
For example, \cite{Lam:2023} analyzed HST WFC3 UVIS F814W imaging of the Bulge and found that a faint star 10 pixels away from a star 3 magnitudes brighter will bias the faint star's measured position by 2 mas.

Crowding can potentially be evaluated with \emph{JWST}, as its larger aperture enables higher angular resolution to provide a ground truth for de-blending efforts.
Ground-based facilities equipped with adaptive optics may also help, although the fields of view are even smaller than JWST (e.g., Keck/OSIRIS ($20'' \times 20''$), Subaru/IRCS ($54'' \times 54''$), VLT/ERIS/NIX ($55'' \times 55''$)).

\emph{Saturation.}
Although this affects a small fraction of sources, saturation has a large impact on characterizing the orbits of unresolved binaries.
Closer stars (which will have larger projected orbits) will tend to be brighter, and being brighter enables better astrometric precision; these are thus the best candidates for characterizing binaries.
There are several proposed methods to extract precise astrometry from saturated stars by using the outer Airy disk pixels or using the diffraction spikes \citep{Gould:2015, Melchior:2018}.

\emph{Chromatic effects.} 
The F146 filter is very wide (1 $\mu$m) and thus chromatic effects could be an issue because stars with different colors will have different PSFs.
The observations taken with the F087 and F213 filters can help characterize the colors of the stars, enabling accurate modeling of the PSFs and mitigating the impact on the astrometric precision.
We note that Gaia's $G$-band filter is also very wide and its astrometry is color-dependent; chromatic effects introduce strong systematic effects that must be measured and calibrated out \citep{Lindegren:2021}.

\emph{Geometric distortion solutions.}
Simulations of the planned LMC astrometric calibration strategy of the Roman WFI find geometric distortion can be corrected to the Poisson noise limit, in the absence of systematics \citep{Bellini:2024}.
However, there will certainly be systematics, due to electronics (e.g., readout hysteresis), detectors (e.g., manufacturing defects), and the spacecraft itself (e.g., out-gassing which changes the plate scale); some of these systematics are time dependent.
However, it is expected that distortion should be correctable at the few $\times 0.1\%$ of a pixel \citep{Bellini:2024} which is better than our assumed astrometric precision floor.

\emph{Brighter-fatter effect.}
Bright sources imaged with HxRG detectors are larger than faint sources because as charge accumulates within a pixel, the substrate voltage changes, causing the pixel's depletion region to contract; this causes charge to be redistributed away from the brightest central pixels, ``fattening" the bright source \citep{Plazas:2018}.
If not properly treated, this will result in most faint stars being measured less precisely than the bright stars used to construct the PSF models.
We note that precise characterization and correction of the brighter-fatter effect is critical for \emph{Roman} to meet its weak lensing cosmology science requirements.

\subsection{Comparison to other works}

\subsubsection{\cite{Gandhi:2023}}

In a white paper submitted as part of the \emph{Roman} core community survey definition process, \cite{Gandhi:2023} proposed searching for non-interacting NS and BH binaries with \emph{Roman} GBTDS astrometry.
However, the discussion on the number of detections was qualitative and did not make any predictions; we thus cannot make comparisons to this work.

\subsubsection{\cite{Eastep:2026}}

\cite{Eastep:2026} independently conducted a study similar to this work.
They perform Galactic population synthesis with a focus on quiescent binary systems containing compact object companions, and assess their astrometric detectability with \emph{Roman} and \emph{Gaia}.
For the population synthesis, they use stellar models from the Binary Population and Spectral Synthesis (BPASS) code, and the Galactic model is from the Feedback In Realistic Environments (FIRE) simulations. 
\cite{Eastep:2026} predict a total of $8.9 \times 10^4$ detached BH+star and $1.8 \times 10^6$ detached NS+star binaries across the entire Galaxy, where $1.0 \times 10^2$ BH+star and $4.2 \times 10^3$ NS+star binaries are within the GBTDS fields; these numbers were calculated by summing the total number of pre-interaction and post-interaction binaries in their Table 1 and share via private communication.
In comparison, our population synthesis produces $4.4 \times 10^3$ NS+star and $3.7 \times 10^2$ BH+star detached binaries.
We note this comparison is not completely equivalent, as \cite{Eastep:2026} simulate the Galactic Center field, which we do not; in addition, \cite{Eastep:2026} model systems down to F$146_{\textrm{AB}} = 30$ mag while we only include systems down to F$146_{\textrm{Vega}} = 23.5$ mag.
However, we do not expect this to change the number of NS+star and BH+star binaries by more than a factor of a few.
This suggests the underlying NS+star and BH+star population synthesis in this work and \cite{Eastep:2026} result in a similar number of systems, to order of magnitude.

As in this work, \cite{Eastep:2026} use the \emph{Roman} GBTDS astrometric noise model from \cite{McKinnon:2026}.
However, they assume the best possible case for the achievable per-epoch astrometric precision, corresponding to a centroiding precision of $\approx 0.01\%$ pixel at the bright end.
In addition, they do not simulate or model individual epoch astrometry; their detectability criteria are based on comparing the size of the astrometric wobble $a_0$ to the J2016 position uncertainty $\sigma_{pos}$ from \cite{McKinnon:2026}, and requiring the orbital period to be less than the time baseline of the GBTDS ($P_{orb} \leq 5$ years).
Note that $\sigma_{pos}$ is not the same as the per-exposure astrometric precision $\sigma_{ast}$; $\sigma_{pos}$ for the \emph{Roman} GBTDS is $20 - 40 \times$ smaller than $\sigma_{ast}$.
Their optimistic (pessimistic) prediction defines detectability as $a_0 > \sigma_{pos}$ ($a_0 > 3 \sigma_{pos}$); using this criteria, \cite{Eastep:2026} find $36 \pm 6$ ($21 \pm 5$) BH+star binaries and $603 \pm 50$ ($92 \pm 22$) NS+star binaries will be detectable in the GBTDS.

We apply the selection cuts outlined in Section 2.4.3. of \cite{Eastep:2026} to our simulated population and compare the results.
We find 971 (538) NS+star and 104 (59) BH+star binaries detectable with the \cite{Eastep:2026} optimistic (pessimistic) criteria; this corresponds to 22\% (12\%) of the NS+star and 28\% (16\%) of the BH+star intrinsic population.
The detection criteria of \cite{Eastep:2026} are thus significantly more optimistic than in this work; using their criteria we would increases the number of detected systems by $5 - 10\times$ for BH+star and $40 - 75\times$ for NS+star binaries.

\subsection{False positives and binary contaminants \label{sec:Binary contaminants}}

As described in Section \ref{sec:Identification of binary stars}, binary star candidates are identified by large $z_{1s}$, indicating the data are poorly described by a single star model.
False positives will be introduced by sources with poor single-star fits that are not astrometric binaries.
For a criterion such as $z_{1s}$, single-star statistical false positives (i.e., single sources that happen to be extreme outliers falling in the $5\sigma$ tail) are negligible.
However, the false positive rate due to any kind of instrumental or astrophysical sources is unknown.

False positives have dominated the search for detached BH + star binaries, and will almost certainly continue to do so, given the rarity of detached BH systems. 
For example, in \emph{Gaia} DR2 and DR3, two common sources of large \texttt{ruwe} not attributed to unresolved binaries are photometrically variable sources and marginally resolved binaries\footnote{We note there are also a completely different set of false positives when selecting BH candidates spectroscopically  or photometrically; these are not discussed here.} \citep{Belokurov:2020, El-Badry:2024b}.
The re-normalization procedure used to calculate \texttt{ruwe} is color and magnitude dependent, and assumes the source does not photometrically vary; variable sources violate these assumptions and have elevated \texttt{ruwe}. 
Although marginally resolved binaries are indeed binaries, their \texttt{ruwe} does not map to the expected orbital parameters, because the \texttt{ruwe} measured is due to large jumps in the centroid from the changing scan angle relative to the orientation of the binary.
In the context of detached BH + star searches, hierarchical triples are another false positive \citep{Nagarajan:2025b, Lam:2026, Simon:2026}, as the outer binary produces a large astrometric signal and the inner binary produces a large spectroscopic signal.
Finally, incorrect measurements due to non-astrophysical reasons (e.g., de-icing, micrometeorite hits, detector issues) may also be a source of false positives.

Some of these false positives  may not be as problematic in the \emph{Roman} data.
For example, variable stars should be even more easily identifiable in \emph{Roman} than \emph{Gaia}, due to the very dense time-series observations.
The issue with very wide partially resolved binaries is also \emph{Gaia}-specific, due to its 1-D astrometry, asymmetric PSF, and scanning law; in \emph{Roman}, these marginally resolved binaries would be easily identifiable via a poorly fit single-star PSF.
Hierarchical triples are also primarily an issue for candidates with a massive and hot stellar companion; most of the stars toward the Bulge are from an old, lower-mass population.

The most anticipated astrometric false positive in the \emph{Roman} GBTDS will be blending due to unresolved sources at small ($<10\%$ pixel) separations which give rise to spurious proper motions.
This is the most common false positive in the $\omega$ Cen search for detached BHs (M. Whitaker et al., in preparation).
Given the crowded Bulge, the probability of close blended sources is relatively high.
How easily these false positives can be identified will be a strong function of the separation, contrast, and SNR.
They can be distinguished by color-dependent positions (e.g., F087 vs. F213), or by inspection of PSF fit quality.
Follow-up imaging at higher resolution with JWST or large ground-based observatories equipped with adaptive optics can be used to rule out more difficult cases; having longer time baselines will also help.
The largest uncertainty is the ratio of false to true positives, and how feasible it will be to rule out those false positives.

\subsection{Spectroscopic follow-up}

With \emph{Gaia}, astrometric orbits are confirmed (or ruled out) via dedicated radial velocity follow-up to validate whether the astrometric and spectroscopic measurements are consistent with each other.
Although two BHs were confirmed with this method in Gaia DR3 \citep{El-Badry:2023a, Chakrabarti:2023, El-Badry:2023b}, false positives dominated the candidate list. 
Even with completely mapped spectroscopic orbits, a precise understanding of why specific false positives occur has been elusive \citep{Simon:2026}.

Given a potentially high false positive rate, a major question is how feasible spectroscopic follow-up will be for \emph{Roman}.
As reference, a population of $1 M_\odot$ star + $10 M_\odot$ BH binaries with a flat eccentricity distribution, isotropic orientations, and orbital periods $P_{orb} = 0.3, 1, 3, 10, 30$ year will have RV semi-amplitudes $K \approx 90^{+50}_{-30}, 60^{+30}_{-20}, 40^{+20}_{-15}, 30^{+15}_{-10}, 20^{+10}_{-5}$ km s$^{-1}$, respectively.
With RV uncertainties $\lesssim 20$\% of the RV semi-amplitude, an orbit can be fairly easily confirmed or refuted with sufficient phase coverage.
This implies a desired RV precision of $\lesssim 4$ km s$^{-1}$ per epoch will generally suffice as long as the orbit is not near face on.
In addition to the instrument itself, the RV precision achievable will depend on the spectral type of the star (the number, width, and depth of lines); it is much easier to achieve precise RVs on cool stars with many deep narrow lines than on hot stars with broad shallow lines.

Here, we explore the feasibility of spectroscopic follow-up for BH candidates. 
In addition to confirmation of astrometric orbits, spectroscopy is useful for characterizing the luminous companion, in particular to measure its metallicity and improve its mass measurement.

A major potential issue with spectroscopic follow up will be crowding.
Seeing-limited observations for fainter stars will be difficult.
For example, the median natural seeing is about 0.6'' (FHWM) at Las Campanas Observatory, Chile.
The diffraction-limited FWHM of Roman at F146 is 115 mas.
Thus, the spatial resolution is a factor of about 5 worse on the ground, meaning that the number of sources per beam on the ground will be a factor of about 30 times higher.
Inspecting Figure \ref{fig:luminosity}, there will be 1 source per ground-based seeing beam around F146$_{\textrm{Vega}} \approx 22$ mag.
However, as long as the sources are not completely overlapping, there are likely techniques that will enable these follow up; for example, rotating the slit, getting a spectrum of nearby stars to subtract them out, or using an integral field spectrograph instead of a slit- or fiber-based instrument.
The high-resolution \emph{Roman} observations could also potentially be used to deblend spectra, which has been demonstrated with integral field spectroscopy observations \citep{Kamann:2013}.

Appendix \ref{app:Forecasts for Spectroscopic Follow-up} considers the feasibility of spectroscopic follow up with several specific instruments and facilities.
We find the Roman/Grism observations that are part of the GBTDS will be insufficient.
Follow-up with current 6 -- 10-m ground-based facilities will be technically feasible, but the observations will be fairly expensive (i.e., a few hours per epoch).
In the ELT era, the observational cost will be significantly lower (i.e., less than an hour per epoch).

\section{Conclusions \label{sec:Conclusions}}

We study the prospects of detecting and characterizing unresolved astrometric binaries in the \emph{Roman} GBTDS, with a focus on BH+star binaries. 
We combine population synthesis simulations with a realistic \emph{Roman} noise model to forecast yields for the GBTDS.
We find that if an astrometric precision of 1\% of a pixel can be achieved (comparable to \emph{HST} and \emph{JWST}), $\mathcal{O}(10)$ BH+star and $\mathcal{O}(10)$ NS+star binaries can be detected.
This forecast assumes the occurrence rate of BHs and NSs in AU-scale orbits around stars is $10^{-7}$ and $10^{-6}$, respectively; the number of detectable systems scales linearly with this assumption about the underlying population.
The BH+star binaries will have typical mass measurements uncertainties of $\sim 25\%$.
The major uncertainties in this prediction are the compact object population synthesis assumptions (from which the number of detectable systems directly scales), the achievable long-term astrometric precision of \emph{Roman}, and the number of false positives.
\emph{Roman} astrometry will be key to revealing the large hidden population of stellar-mass BHs, and we advocate for observing strategies, calibrations, pipelines, and data analysis that will enable \emph{Roman} to reach its full astrometric potential.

\begin{acknowledgements}

C.Y.L. acknowledges support from the Harrison and Carnegie Fellowships.
Support for this work was provided by NASA through the NASA Hubble Fellowship grant HST-HF2-51609.001-A awarded by the Space Telescope Science Institute, which is operated by the Association of Universities for Research in Astronomy, Inc., for NASA, under contract NAS5-26555.
N.S.A. acknowledges support from the Heising-Simons Foundation under grant No. 2022-3542 and the H2H8 foundation.
M.W.H. is supported by the Brinson Prize Fellowship.
P.M. acknowledges that this work was performed under the auspices of the U.S. Department of Energy by Lawrence Livermore National Laboratory under Contract DE-AC52-07NA27344. 
The document number is LLNL-JRNL-2024312.
Thanks to Max Mutchler at the Roman Help Desk for answering questions about the GBTDS cadence.
Thanks to Macy Huston, Jay Anderson, and Matthew Whitaker for helpful discussions.
Thanks to Matthew Penny for providing the code to create the \emph{Roman} GBTDS field pointings in Figure 1.
Thanks to Erin Eastep for discussions on \citep{Eastep:2026}.
Thanks to the referee for comments that helped to improve the clarity of the paper.
This research was supported in part by grant NSF PHY-2309135 to the Kavli Institute for Theoretical Physics (KITP).
This research used resources of the National Energy Research Scientific Computing Center (NERSC), a Department of Energy User Facility using NERSC award HEP-ERCAP0030957.

\software{gaia\_roman\_astrometry \citep{kevin_mckinnon_2026_18407082}}

\end{acknowledgements}


\appendix

\section{Assumptions about Roman GBTDS and Gaia DR4 performance \label{app:Assumptions about Roman GBTDS and Gaia DR4 performace}}

The forecasts for the number of stars and Roman's astrometric precision are described in Section \ref{sec:Simulations}.
For Roman, we define an effective epoch as 1 day of GBTDS observations.
For sources with astrometric variability on timescales $\gg 1$ day, observations taken within the timespan of 1 day can effectively be treated as a single epoch with an improvement of $\sqrt{N}$ on the astrometric uncertainties assuming an appropriate dither strategy, where $N$ is the number of observations in 1 day; this has been demonstrated on HST with $N \approx 20$ epochs \citep{Anderson:2000, Hosek:2015}.
For the Roman GBTDS high-cadence seasons, $N \approx 100$; as there are 6 high-cadence seasons of about 70 days each, this results in 420 effective epochs.

For Gaia, the expected along-scan error per CCD for DR4 \citep{Brown:2024} is nearly identical to that from DR3 \citep{Holl:2023}.
Thus, we assume the number of bright sources and their per-measurement noise properties do not change from DR3 to DR4; the only difference is the longer time baseline and number of epochs.
We define the astrometric precision per epoch to be the along-scan error per field of view (FOV) transit.
The typical number of transits per scan is 8--9. Assuming the scans are independent, the error per FOV transit is equal to the error per CCD transit divided by $\sqrt{8}$; \cite{El-Badry:2024b} demonstrate the validity of this assumption.
We define an effective epoch to be the number of visibility periods (defined to be the number of groups of observations separated by at least 4 days), which provides a more reliable estimate for position and proper motion uncertainties than the total number of good along-scan observations \citep{Lindegren_etal:2018}.
We assume the number of Gaia DR4 visibility periods is twice that of Gaia DR3, since the time baseline will roughly double from 34 to 66 months.

\section{Details of the $\omega$ Cen black hole search \label{app:Details of the omega Cen black hole search}}

In order to identify astrometrically variable sources,
\cite{Platais:2024} fit a 4-parameter model (i.e., positions and proper motions, but no parallax) to all sources to search for astrometrically accelerating sources.
They also used a periodogram to search for periodicity in the astrometric residuals to their 4-parameter fit, searching for high-significance periodicity not due to parallax (i.e., having periods of 0.5 or 1 year).
\cite{Whitaker:2026} also took a similar approach by fitting a 4-parameter model to all sources and then used a periodogram search with a false alarm probability threshold to identify candidates.
This is different that the methodology used by \cite{Halbwachs:2023} for Gaia DR3 and what is presented in this paper; we advocate using both methods when analyzing the \emph{Roman} data to search for detached binary systems.

Regarding the nature of the astrometrically accelerating source,
\cite{Whitaker:2026} attribute the difference in interpretation to having observations spanning a longer temporal baseline (23 vs. 13 years) that provided good coverage of the orbit's pericenter passage, and the modeling algorithm of \cite{Platais:2024} disfavoring longer period solutions. 
The orbital period and photocenter orbit semimajor axis measured presented in  \cite{Whitaker:2026} are highly uncertain ($94^{+63}_{-42}$ year and $31^{+15}_{-12}$ AU); despite this, the mass of the BH is well-constrained to $4.46^{+1.22}_{-1.01} M_\odot$.
This is because the acceleration is proportional to the mass, which in turn is proportional to $a_0^3/P_{orb}^2$ according to Kepler's law; thus, even if $P_{orb}$ and $a_0$ are individually poorly constrained, their ratio is well-measured in cases where an acceleration is detected \citep{Lucy:2014}.

\section{Forecasts for Spectroscopic Follow-up \label{app:Forecasts for Spectroscopic Follow-up}}

\emph{Roman/Grism.} 
There will be 30 slitless grism ``spectroscopic snapshot" observations taken as part of the GBTDS.
Due to crowding, only the brightest sources $K_s \lesssim 14$ mag will have RV measurements with a precision of $\sim 10$ km s$^{-1}$ \citep{ROTAC:2025}; however, these sources will be saturated in the Bulge imaging, and the fainter sources with good astrometry will be unlikely to have useful RV measurements.

\emph{VLT/X-Shooter.}
\cite{Heida:2015} observe a red supergiant with X-Shooter, using the 0.6'' slit for the NIR arm ($R \approx 8000$) and achieve a SNR of $\approx 10$ per resolution element in J and H band. 
They measure radial velocities by cross-correlating against a template, and achieve a statistical error of 1 km s$^{-1}$.
The RV precision is dominated by the X-Shooter wavelength calibration (2 km s$^{-1}$) and the RV uncertainty of the template itself (3 km s$^{-1}$).
According to the X-Shooter exposure time calculator\footnote{https://www.eso.org/observing/etc/}, SNR of 10 in H-band can be achieved for a K7V star spectrum of H=19 Vega in default sky conditions with a 1.5 hour exposure.

\emph{Magellan/FIRE.}
Scaling off the setup described in \cite{Simcoe:2012}, for the 0.6'' slit ($R=6000$) for $H \approx 19$ mag (Vega), a SNR $\approx 10$ per resolution element can be achieved with a 3.5 hour exposure in median conditions.

\emph{Keck/OSIRIS.}
According to the OSIRIS exposure time calculator\footnote{https://oirlab.ucsd.edu/osiris/etc/}, for a 1.5 hr exposure of $H_{bb} = 19$ mag (Vega) for a solar-type star spectrum, 35 mas plate scale, Strehl ratio 0.5, and Fried parameter 40 cm, the SNR is about 10 per spectral flux element for an $R \sim 4000$ spectrum.
This assumes OSIRIS is being used with the adaptive optics (AO) system; the SNR is highly dependent on the performance of the AO system, in addition to the observing conditions.

\emph{ELT.}
According to the ELT exposure time calculator, assuming a 39m telescope, for a K7V spectrum star of H=20 Vega, default sky conditions, using LT/MCAO, a 1 hour exposure will result in a SNR of 35 at H-band per resolution element  for $R=10000$.
Thus, follow-up of these sources will be easy in the age of ELTs.

\bibliography{sample701}{}
\bibliographystyle{aasjournalv7}

\end{document}